\documentclass[sigconf, nonacm]{acmart}

\usepackage[ruled,linesnumbered,vlined]{algorithm2e}
\usepackage{amsmath}
\usepackage{amsthm}
\usepackage{array}
\usepackage{algorithmic}
\usepackage{bm}
\usepackage{multirow} 
\usepackage{appendix}
\usepackage{xcolor}
\usepackage{enumitem}
\usepackage{soul}
\usepackage{balance}
\usepackage[compact]{titlesec}
\usepackage{framed}
\usepackage{xcolor}
\titlespacing{\section}{0pt}{0.8ex}{0ex}      
\titlespacing{\subsection}{0pt}{0.8ex}{0ex}
\titlespacing{\subsubsection}{0pt}{0.8ex}{0ex}
\usepackage{amsthm}
\newtheoremstyle{tight}  
  {1pt}   
  {1pt}   
  {\itshape}  
  {}      
  {\bfseries} 
  {.}     
  { }     
  {}      

\theoremstyle{tight}
\newtheorem{theorem}{Theorem}
\newtheorem{lemma}{Lemma}
\newtheorem{definition}{Definition}
\newtheorem{remark}{Remark}
\newtheorem{example}{Example}
\makeatletter
\renewenvironment{proof}[1][\proofname]{%
  \par\vspace{2pt}
  \pushQED{\qed}\normalfont
  \topsep2pt \partopsep0pt \itemsep0pt \parsep0pt
  \trivlist
  \item[\hskip\labelsep\bfseries #1\@addpunct{.}]%
}{%
  \popQED\endtrivlist\@endpefalse
  \vspace{2pt}
}
\makeatother

\newcommand\vldbdoi{XX.XX/XXX.XX}
\newcommand\vldbpages{XXX-XXX}
\newcommand\vldbvolume{19}
\newcommand\vldbissue{4}
\newcommand\vldbyear{2025}
\newcommand\vldbauthors{\authors}
\newcommand\vldbtitle{\shorttitle} 
\newcommand\vldbavailabilityurl{URL_TO_YOUR_ARTIFACTS}
\newcommand\vldbpagestyle{empty} 

\newtheorem{property}{Property}

\newcommand{\stitle}[1]{\noindent{{\bf #1}}}
\newcommand{\sstitle}[1]{\noindent{\underline{#1}}}
\newcommand{\ssstitle}[1]{\noindent{\textit{#1}}}

\definecolor{shadecolor}{rgb}{0.9,0.9,0.9}

\SetAlgoNoEnd

\begin{document}
\title{Efficient Partition-based Approaches for Diversified Top-$k$ Subgraph Matching}
%
\author{Liuyi Chen}
\affiliation{%
  \institution{Hunan University}
  \city{Changsha}
  \state{China}
}
\email{liuyi.chen@hnu.edu.cn}

\author{Yuchen Hu}
\affiliation{%
  \institution{Hunan University}
  \city{Changsha}
  \state{China}
}
\email{shadowhyc@hnu.edu.cn}

\author{Zhengyi Yang}\thanks{* Zhengyi Yang and Xu Zhou are the corresponding authors.}
\affiliation{%
  \institution{The University of New South Wales}
  \city{Sydney}
  \country{Australia}
}
\email{zhengyi.yang@unsw.edu.au}

\author{Xu Zhou}
\affiliation{%
  \institution{Hunan University}
  \city{Changsha}
  \state{China}
}
\email{zhxu@hnu.edu.cn}

\author{Wenjie Zhang}
\affiliation{%
  \institution{The University of New South Wales}
  \city{Sydney}
  \country{Australia}
}
\email{wenjie.zhang@unsw.edu.au}

\author{Kenli Li}
\affiliation{%
  \institution{Hunan University}
  \city{Changsha}
  \state{China}
}
\email{lkl@hnu.edu.cn}

\begin{abstract}
Subgraph matching is a core task in graph analytics, widely used in domains such as biology, finance, and social networks. Existing top-$k$ diversified methods typically focus on maximizing vertex coverage, but often return results in the same region, limiting topological diversity. 
We propose the Distance-Diversified Top-$k$ Subgraph Matching (DT$k$SM) problem, which selects $k$ isomorphic matches with maximal pairwise topological distances to better capture global graph structure. To address its computational challenges, we introduce the Partition-based Distance Diversity (PDD) framework, which partitions the graph and retrieves diverse matches from distant regions.
To enhance efficiency, we develop two optimizations: embedding-driven partition filtering and densest-based partition selection over a Partition Adjacency Graph. Experiments on 12 real world datasets show our approach achieves up to four orders of magnitude speedup over baselines, with 95\% of results reaching 80\% of optimal distance diversity and 100\% coverage diversity.
\end{abstract}
\maketitle
\pagestyle{\vldbpagestyle}
\begingroup\small\noindent\raggedright\textbf{PVLDB Reference Format:}\\
\vldbauthors. \vldbtitle. PVLDB, \vldbvolume(\vldbissue): \vldbpages, \vldbyear.\\
\href{https://doi.org/\vldbdoi}{doi:\vldbdoi}
\endgroup
\begingroup
\renewcommand\thefootnote{}\footnote{\noindent
This work is licensed under the Creative Commons BY-NC-ND 4.0 International License. Visit \url{https://creativecommons.org/licenses/by-nc-nd/4.0/} to view a copy of this license. For any use beyond those covered by this license, obtain permission by emailing \href{mailto:info@vldb.org}{info@vldb.org}. Copyright is held by the owner/author(s). Publication rights licensed to the VLDB Endowment. \\
\raggedright Proceedings of the VLDB Endowment, Vol. \vldbvolume, No. \vldbissue\ %
ISSN 2150-8097. \\
\href{https://doi.org/\vldbdoi}{doi:\vldbdoi} \\
}\addtocounter{footnote}{-1}\endgroup
\ifdefempty{\vldbavailabilityurl}{}{

\begingroup\small\noindent\raggedright\textbf{PVLDB Artifact Availability:}\\
The source code, data, and/or other artifacts have been made available at \url{https://github.com/Twilight-Shadow/DiversifiedTopkSubgraphMatching}.
\endgroup
}
\section{Introduction}
Subgraph matching is one of the most fundamental tasks in graph analysis. Given a query graph and a data graph, it finds all subgraphs from the data graph that are isomorphic to the query graph. It serves as an effective means to extract valuable information from complex networks, supporting applications such as pattern recognition \cite{GE2025110797,hao2019patmat}, query optimization \cite{qo1,qo2,lai2019improving,ding2024fgaq,chen2025accelerating}, and anomaly detection \cite{ad1}.

Numerous efforts \cite{first,cfl,gql,64,vf2+,turboiso,gup,ceci,7907163,dualsim,74,77,81,yang2023hgmatch} have been devoted to addressing this problem in the literature.
However, as the scale of data graphs grows, sometimes reaching billion-scale, two key issues arise: 1) Subgraph matching is NP-hard, making it computationally challenging to enumerate all possible matches. 2) The number of resulting subgraphs can be overwhelmingly numerous \cite{2,huge,ptab,tkzhang,tkzou2,7}, making them difficult to be analyzed effectively.
A common practice in the literature is to enumerate the first $k$ matches \cite{qin2012,6816703,doi:10.1142/S0218001418500209,10.1145/3132847.3132966,10.1007/978-3-642-31235-9_14}, where $k$ is a user-defined constant that limits the maximum number of results. While this strategy reduces computational overhead, the returned matches are often redundant or overly similar. To address this, diversified top-$k$ subgraph matching has been studied \cite{74, 77, 81}, which aims to return a set of $k$ subgraphs that are not only isomorphic to the query but also diverse in structure or content, thereby enhancing interpretability and utility in downstream applications. 

\stitle{Motivation.} 
Diversity metrics play a key role in top-$k$ query processing by assessing the quality of returned results, aiming to reduce redundancy and enhance representativeness~\cite{76, 10.1145/3588924}. This concept naturally extends to top-$k$ subgraph matching.
Existing works~\cite{73,81} primarily aim to enhance result diversity by maximizing \textit{vertex coverage}, which measures the number of distinct vertices covered by the selected set of subgraph matches. 
%
However, vertex coverage alone may fail to capture the true \textit{structural dispersion} of subgraph matches, as it overlooks the \textit{topological proximity} among the selected results in the data graph. That is, two subgraph matches may collectively cover many unique vertices, yet still be located in close proximity to each other, possibly within the same neighbourhood. In such cases, the result set may exhibit limited structural diversity, offering redundant or localized views.

\stitle{Distance-based Diversity.}
To address this limitation, we introduce the problem of \textit{Distance-Diversified Top-$k$ Subgraph Matching} (DT$k$SM), which retrieves $k$ subgraph matches that are not only isomorphic to the query graph but also maximally dispersed throughout the data graph. This formulation explicitly promotes diversity by maximizing the pairwise topological \textit{distances} among the selected matches in the data graph, resulting in a more representative and interpretable answer set that better captures the broader structural landscape of the graph.

%

\begin{figure}[t]
    \centering
    \includegraphics[width=0.45\textwidth]{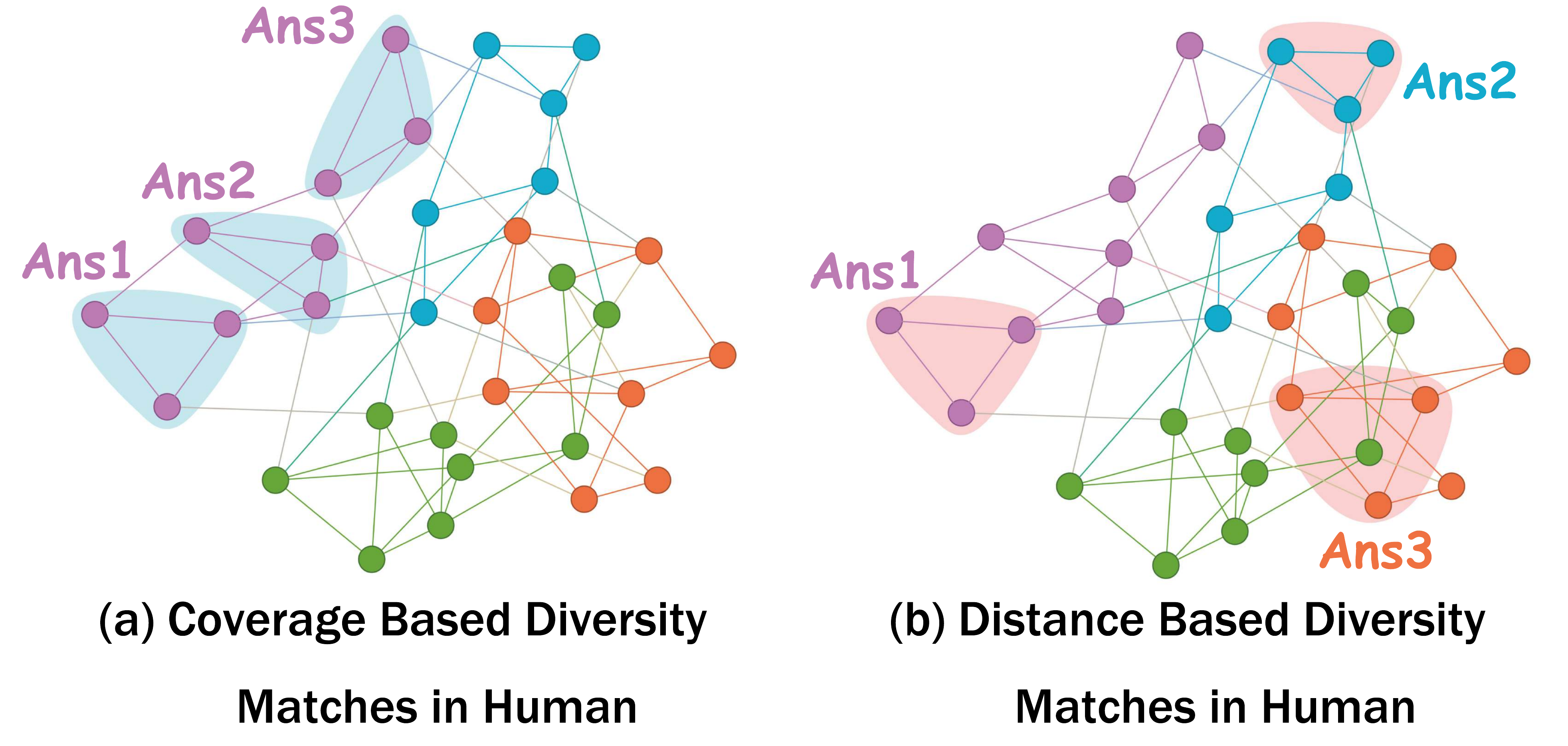}
    \caption{Example of Top-$3$ diversified subgraph matches.}
    \label{fig:enter-label_1}
\end{figure}

\begin{example}

Figure~\ref{fig:enter-label_1} shows the top-$k$ subgraph matching results on  real-world dataset \textit{Human}.
%
When using vertex coverage, as shown in Figures~\ref{fig:enter-label_1}(a), the selected matches (highlighted in blue) are concentrated in specific local regions. Although these matches exhibit minimal node overlap, their spatial proximity limits topological diversity, reducing their ability to capture the global structural characteristics of the data graph.
In contrast, Figures~\ref{fig:enter-label_1}(b) show the results using distance-based diversification. The selected matches (highlighted in red) are more widely distributed across the graph. This dispersion results in better diversity, allowing the results to more effectively represent the structural heterogeneity of the graph.
\end{example}

\stitle{Applications.} 
Beyond the motivating example, distance-based diversity has broad utility across domains.

\sstitle{Biomedical Interaction Networks.}
Subgraph matching supports disease pattern discovery in biomedical networks, such as protein–protein interaction graphs~\cite{Shen2014,agrawal2017,klughammer2024mapping}. In cancer research, a query graph may encode a metastasis-related structure, and its top-$k$ matches reveal potential pathways or organ sites of spread~\cite{bio1,agrawal2017}. However, coverage-based methods often yield results clustered in a single region, while metastasis typically spans multiple distant organs~\cite{bio2,Su2016}. Distance-based diversity enables spatially dispersed matches (e.g., lungs, liver, brain), uncovering independent metastatic routes and aiding prognosis, treatment planning, and biomarker discovery~\cite{bio3,xu2025}.

\sstitle{Financial Transaction Networks.}
Subgraph matching detects recurring suspicious structures such as fraud rings and money laundering schemes~\cite{zou24,fin01}. Coverage-based diversification tends to cluster results within one institution or group, missing distributed activity. In contrast, DT$k$SM promotes dispersion across institutions, user groups, or regions, improving both coverage and robustness in financial crime detection~\cite{egressy2024,fin02}.

\sstitle{Social Networks.}
Subgraph matching helps identify patterns like influencer–follower motifs or tightly-knit triads~\cite{mesnards2018,jiang2020user}. Coverage-based methods often concentrate results in one community, limiting the detection of replicated behaviors such as misinformation campaigns~\cite{sharma2021,zhang2015}. Distance-based diversification instead selects dispersed matches across distinct communities, enhancing the detection of distributed influence operations and supporting opinion analysis, anomaly detection, and moderation~\cite{pacheco2021,Ma2022,chen2012clustering}.

\stitle{Challenges.}
While distance-based diversified top-$k$ subgraph matching yields more insightful and globally representative results, computing such results is non-trivial. Unlike coverage-based diversification, which can often be addressed through local reasoning (e.g., avoiding vertex overlap), distance-based diversification requires a global view of the graph to assess topological dispersion. Identifying the optimal result set necessitates enumerating all subgraph matches, as pairwise distances must be compared across the entire result set. This amplifies the computational complexity, particularly in large-scale graphs where both subgraph enumeration and distance evaluation are resource-intensive.
The primary challenges arise from two sources:

\ssstitle{Subgraph Matching.} 
Subgraph matching is inherently NP-hard. This intrinsic complexity presents a fundamental bottleneck, especially when applied to large data graphs or complex query structures.

\ssstitle{Distance Computation.} 
Ensuring structural diversity involves computing pairwise distances among a potentially large number of candidate matches. As the candidate space grows, distance evaluation becomes increasingly costly, particularly in large-scale graphs where such computations demand significant resources.

\stitle{Research Question.} This leads to the critical research question: \ul{\textit{how to efficiently find high-quality distance-diversified subgraph matching results while minimizing computational overhead?}}





\stitle{Contributions.}
In this paper, we study the problem of \textit{distance-diversified top-$k$ subgraph matching}, which seeks $k$ isomorphic matches of a query graph in a data graph such that pairwise topological distances are maximized. To efficiently retrieve such results, we propose the \textit{Partition-based Distance Diversity (PDD)} framework, which partitions the data graph and retrieves matches from topologically distant partitions. To further improve efficiency, we introduce two optimizations: \textit{embedding-driven partition filtering}, which uses node embeddings to prioritize partitions likely containing matches, and \textit{density-based partition selection}, which models the choice of partitions as a densest subgraph problem over the \textit{Partition Adjacency Graph} (PAG), ensuring selected partitions are both dispersed and structurally rich.
The key contributions of this work are:

\begin{itemize}[leftmargin=4pt, nosep]
   \item \textbf{Novel Problem Formulation.}
   We define the \emph{Distance-Diversified Top-$k$ Subgraph Matching (DT$k$SM)} problem, which aims to select $k$ isomorphic subgraph matches that maximize pairwise topological distances. 
   This formulation addresses the limitations of existing coverage-based diversified top-$k$ subgraph matching, which typically avoid overlap but often return results within the same local region, thereby failing to capture the global structural characteristics of the data graph.
   We further prove that: 
   1) solutions to DT$k$SM also maximize vertex coverage as a natural consequence of structural dispersion, and 
   2) the DT$k$SM problem remains NP-hard, even when all subgraph matches are pre-computed.
   
   \item \textbf{Partition-Based Distance Diversity Framework.} 
   To avoid exhaustive enumeration of all matches and the high cost of pairwise distance computations, we propose the \emph{Partition-based Distance Diversity (PDD)} framework. This framework first partitions the data graph and precomputes the distances between partitions. It then retrieves subgraph matches from both intra-partition and inter-partition regions that are topologically distant from one another.
   This strategy significantly reduces the search space and computational overhead while effectively preserving the distance-based diversity objective of DT$k$SM. Moreover, the partitioned structure naturally enables parallel execution of the matching process across different graph regions, further improving scalability. 
    
    \item \textbf{Efficient Partition Filtering and Selection.} 
    To further improve performance, we introduce two optimizations: \emph{embedding-driven partition filtering} and \emph{densest-based partition selection}.
    The filtering phase leverages learned embeddings to estimate which partitions are likely to contain valid subgraph matches, enabling the system to filter out unpromising partitions. This reduces unnecessary search and significantly lowers latency.
    The selection phase reformulates the task of choosing $k$ well-separated partitions as a densest subgraph problem over a \emph{Partition Distance Graph (PDG)}, where edges connect non-adjacent partitions and are weighted by inter-partition distances. This formulation enables efficient selection of topologically dispersed partitions, thereby promoting structural diversity while maintaining computational efficiency.
    
   \item \textbf{Comprehensive Experimental Evaluation.} 
   We conducted extensive experiments on 12 datasets, including both labeled and unlabeled graphs. The results demonstrate that our method achieves up to \textbf{four orders of magnitude} speedup over existing baselines and \textbf{one order of magnitude} on average. 
   In terms of effectiveness, \textbf{95\%} of our results achieve at least \textbf{80\%} of the optimal distance diversity and consistently exhibit optimal coverage diversity. In contrast, only \textbf{3\%} of cases from competing methods reach comparable quality.

\end{itemize}


\section{PRELIMINARIES}\label{sec:2}
In this section, we present the key definitions and propose the problem statement of this paper. 

\subsection{Key Concepts}
We study an undirected and unweighted graph \( G = (V, E) \), where \( V \) denotes the vertex set and \( E \subseteq \{\{u, v\} \mid u, v \in V, u \ne v\} \) is the edge set. Each vertex \( v \in V \) is associated with a categorical label \( \ell(v) \). For a vertex \( v \), the \emph{degree}, denoted as \( \deg(v) \), is defined as the number of edges incident to \( v \). Two vertices \( u \) and \( v \) are said to be \emph{adjacent} if \( \{u, v\} \in E \), i.e., if they are connected by an edge.

\begin{definition}[Shortest Path]
Given a graph $G{=}(V,E)$, the shortest path between two vertices $u,v \in V$ is the path with the minimum total edge weight among all possible paths from $u$ to $v$ in $G$.
\end{definition}

\begin{definition}[Subgraph Isomorphism]\label{def:subgraph_iso}

Given a query graph \( Q = (V_Q, E_Q, \ell_Q) \) and a data graph \( G = (V_G, E_G, \ell_G) \), a subgraph \( M \subseteq G \) is said to be \emph{isomorphic} to \( Q \) if there exists an injective mapping \( f: V_Q \rightarrow V_G \) such that \( \ell_Q(u) = \ell_G(f(u)) \) for all \( u \in V_Q \), and \( (f(u_1), f(u_2)) \in E_G \) for all \( (u_1, u_2) \in E_Q \). The subgraph \( M \), induced by the vertex set \( f(V_Q) \), is referred to as a \emph{match} of \( Q \) in \( G \), and \( f \) is called a \emph{subgraph isomorphism mapping}.
\end{definition}
\begin{definition}[Subgraph Matching]\label{def:subgraph_match}

The goal of \emph{subgraph matching} is to enumerate the complete set \( \mathcal{M} = \{M_1, M_2, \dots, M_t\} \) of subgraphs in \( G \), where each \( M_i \) is isomorphic to \( Q \). The set \( \mathcal{M} \) is referred to as the \emph{match set} of \( Q \) in \( G \).
\end{definition}
\begin{definition}[Diversified Top-$k$ Subgraph matching]\label{def:div_topk}

Given a query graph \( Q \), a data graph \( G \), and an integer \( k \), let \( \mathcal{M} \) denote the match set of \( Q \) in \( G \) as defined in Definition~\ref{def:subgraph_match}. The \emph{diversified top-$k$ subgraph matching} problem aims to find a subset \( \mathcal{R} \subseteq \mathcal{M} \) with \( |\mathcal{R}| = k \) that maximizes a given diversity function \( \mathcal{F}(\mathcal{R}) \).
\end{definition}
\begin{definition}[Coverage-based Diversity \cite{73}]\label{cover dist}
Given a set of $k$ subgraph matches $\mathcal{R} = \{M_1, M_2, \ldots, M_k\}$ in a data graph $G$, the \emph{coverage-based diversity function} is defined as 
\begin{equation}
    \setlength\abovedisplayskip{0pt}
    \setlength\belowdisplayskip{0pt}
    \mathcal{F}_{\mathrm{cov}}(\mathcal{R}) = \left| \bigcup_{i=1}^{k} V(M_i) \right|.
\end{equation}
\end{definition}

However, as illustrated in the introduction, the above definition fails to fully capture the essence of diversity. To address this limitation, we introduce a more comprehensive metric that emphasizes distance-based diversity among the results.

\begin{definition}[Subgraph Distance]\label{def:subgraph_distance}

Given two subgraphs $M_i$ and $M_j$ in a data graph $G$, their distance is defined as:
$d(M_i, M_j) = \mathbf{min}_{u \in M_i,\ v \in M_j} \text{dist}_G(u, v).$
\end{definition}

\begin{definition}[Distance-based Diversity]\label{def:distance}

Given a set of $k$ subgraph matches $\mathcal{R} = \{M_1, M_2, \ldots, M_k\}$ in a data graph $G$, \emph{distance-based diversity} is defined as:
\begin{equation}
    \setlength\abovedisplayskip{0pt}
    \setlength\belowdisplayskip{0pt}
    \mathcal{F}_{\mathrm{dis}}(\mathcal{R})= \max_{i \ne j} \min d(M_i, M_j),
\end{equation}
where $M_i, M_j \in \mathcal{R}$ denote distinct subgraph matches.
\end{definition}

\begin{figure}[t]
    \centering
    \includegraphics[width=0.3\textwidth]{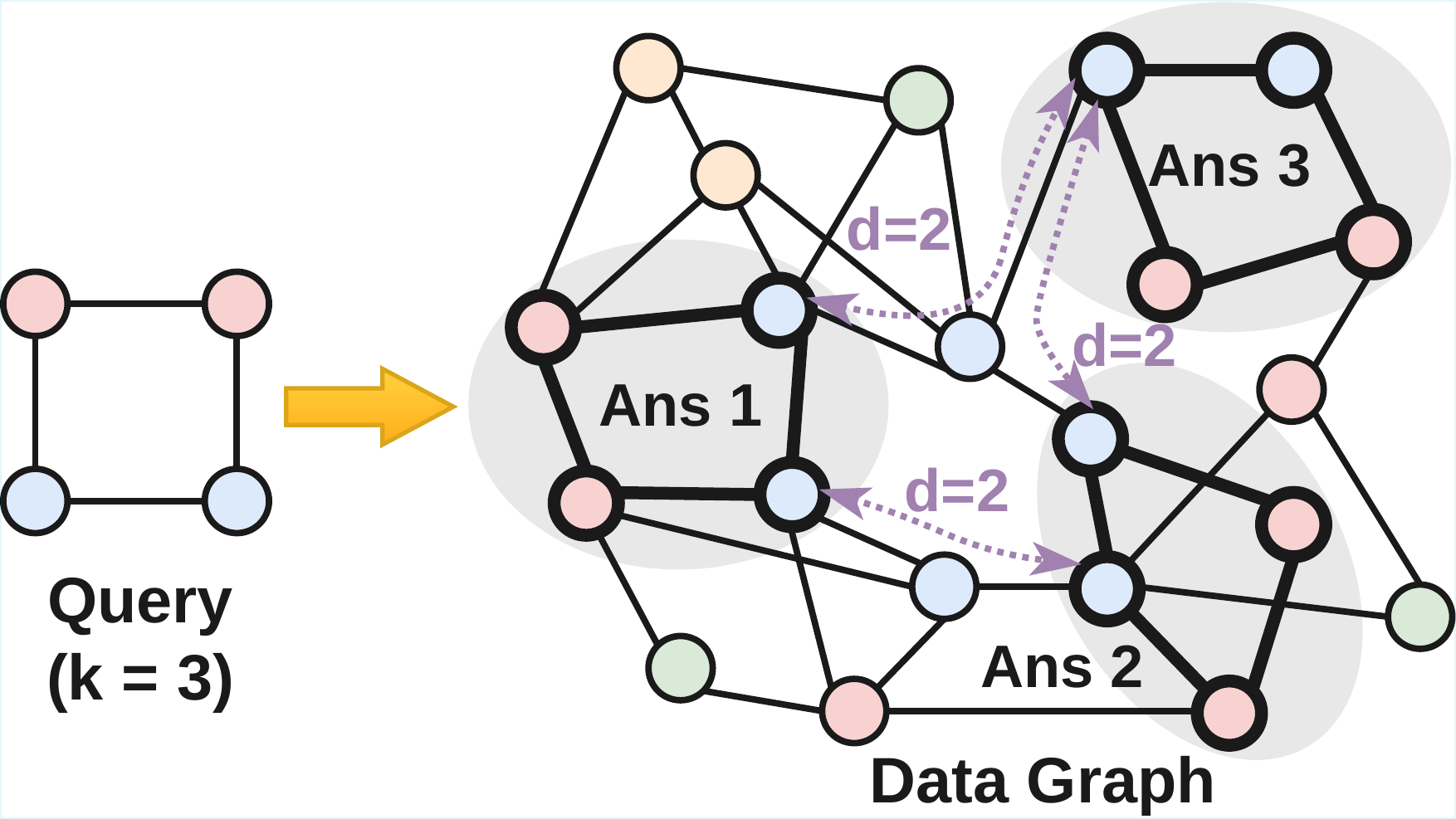}
    \caption{Example of calculating subgraph distance.}
    \label{fig:casestudy}
\end{figure}

\begin{example}
    Figure \ref{fig:casestudy} shows an example of top-$3$ diversified subgraph matches of a simple query. For \(\mathcal{R}=\{Ans_1,Ans_2,Ans_3\}\), \(d(Ans_1,Ans_2)=d(Ans_1,Ans_3)=d(Ans_2,Ans3)=2\), therefore we maximize \(\mathcal{F}_{dis}(\mathcal{R})\) to 2. It can be proven that this solution is optimal in this scenario.
\end{example}

\begin{remark}
The proposed definition of distance-based diversity is theoretically grounded and consistent with established practices in top‑$k$ recommendation systems, where diversity is commonly promoted by maximizing the minimum pairwise distance among items. Similarly, our formulation adopts this principle by maximizing the minimum topological distance between matches, making the metric both intuitive and well-supported by prior theoretical frameworks.
\end{remark}

\subsection{Problem Statement}
\begin{definition}[Distance-Diversified Top-$k$ Subgraph Matching (DT$k$SM)]

    Given a query graph Q and a data graph G, the goal is to find a set of $k$ subgraph matches such that each match is isomorphic to Q, and the result set maximizes the distance-based diversity objective defined in Definition~\ref{def:distance}.
\end{definition}
\begin{remark}
The problem of DT$k$SM can be naturally decomposed into two stages: (1) enumerating all subgraph matches that are isomorphic to the query graph $Q$, and (2) selecting a subset of $k$ matches that maximizes the distance-based diversity objective. 
\end{remark}
\begin{theorem}
Finding the top-$k$ subgraph matches that maximize distance-based diversity is NP-Hard, even when all subgraph matches are pre-computed.
\end{theorem}

\begin{proof}
\noindent Due to page limitations, the detailed proof is provided in the Appendix A (Theorem 1).    
\end{proof}
\section{The Proposed Method} \label{sec:3}
Although exhaustively enumerating the top-$k$ most diverse subgraph matches guarantees optimality, it is computationally infeasible due to the exponential number of isomorphic candidates in large graphs. To overcome this, we introduce the PDD framework that first selects the $k$ most dispersed regions (graphs) via coarsen distance, and then performs subgraph matching in parallel within them. This approach significantly reduces the computational cost.
\begin{figure*}
\vspace{-16pt}
    \centering  \includegraphics[width=16cm,height=4.4cm]{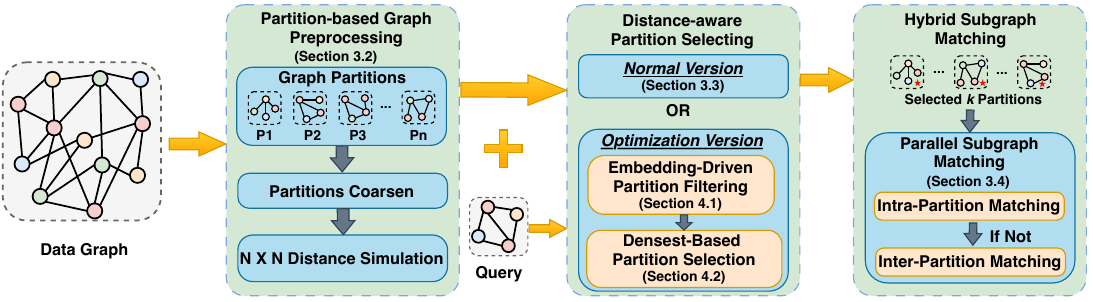}
    \caption{Overview Framework}
    \label{fig:enter-label_3}
\end{figure*}
\subsection{Overview Framework}\label{AA}
The new algorithmic framework is shown in Algorithm \ref{alg:partition_matching}.
The input graph G is divided into $p$ disjoint partitions $\{G_1,\dots, G_p\}$. To guide partition selection, a $p \times p$ distance matrix $D$ for partitions is computed, where each entry $D_{i,j}$ represents the distance between partitions $G_i$ and $G_j$ (will be defined later).
To ensure the diversity among the matches, the algorithm incrementally selects a subset $\mathcal{P}$ from the partition set. Starting from a random partition, each subsequent partition is selected to be the one which owns the maximum distance from the already selected ones, based on matrix $D$. This helps to spread the search across graph regions that are topologically far apart.
Each selected partition $G_i \in \mathcal{P}$ is processed in parallel to search for occurrences of the query subgraph Q. If a match is found within a single partition, it is added to the match set $\mathcal{M}$. If the number of matched subgraphs is still less than $k$, an additional inter-partition matching phase is invoked, where the algorithm searches for matches that span multiple partitions in $\mathcal{P}$.
Finally, the algorithm returns up to $k$-matched subgraphs that are expected to be both correct and topologically diverse.

\begin{algorithm}[t]
\small
\caption{The PDD Framework}
\label{alg:partition_matching}
\SetAlgoLined
\SetAlgoNoEnd
\KwIn{Data graph $G$, query graph $Q$, target size $k$}
\KwOut{Set $\mathcal{M}$ of $k$ diverse subgraph matches}
\texttt{//Partition-based Graph Preprocessing}\\
Partition $G$ into $\{G_1, G_2, \dots, G_n\}$\;
Compute $n \times n$ distance matrix $D$, where $D_{i,j}$ is the distance in the Partition Adjacency Graph (PAG)\;
\texttt{// Distance-aware Partition Selection}\\
Initialize candidate list $\mathcal{P} \gets \emptyset$\;
Randomly select a partition $G_r$ and add to $\mathcal{P}$\;
\While{$|\mathcal{P}| < k$}{
    Identify $G_j$ as the farthest valid neighbor from current $\mathcal{P}$ using $D$\;
    $\mathcal{P} \gets \mathcal{P} \cup \{G_j\}$\;
}
\texttt{// Hybrid Subgraph Matching}\\
Initialize match set $\mathcal{M} \gets \emptyset$\;
\ForEach{$G_i \in \mathcal{P}$ \textbf{in parallel}}{
$\mathcal{M} \gets \mathsf{HySM}(G_i,Q)$ (Algorithm 2)
}
\texttt{// Backtracking (if $|\mathcal{M}| < k$)}\;
\If{$|\mathcal{M}| < k$}{
    Let $m = $k$ - |\mathcal{M}|$\;
    Select $m$ most promising partitions not in $\mathcal{P}$\;
    Perform matching in these partitions to complete $\mathcal{M}$\;
}
\Return{$\mathcal{M}$}
\end{algorithm} 

\stitle{Time Complexity Analysis:} Excluding offline preprocessing (e.g., graph partitioning and distance computation), the online algorithm consists of two phases. In the first phase, selecting $k$ dispersed partitions from $n$ candidates requires $O(nk)$ time, as each round evaluates up to $n$ neighbors. In the second phase, fine-grained subgraph matching is performed parallel on the $k$ selected partitions. Letting $T_\mathrm{match}$ denote the total matching time, the overall online complexity is $O(nk + T_\mathrm{match})$.

\subsection{Partition-based Graph Preprocessing}
\noindent\textbf{Graph Partition.}
Partitioning the graph supports scalability by dividing a large graph into smaller, manageable subgraphs that enable parallel processing and reduce memory overhead. It also facilitates efficient candidate filtering by localizing the search space.
We adopt Distributed-NE \cite{48}, a parallel edge partitioning strategy that grows each partition from a seed vertex via greedy edge expansion. To maintain logical consistency, the scheme uses vertex replication: when an edge spans multiple partitions, its endpoints are duplicated across them. Thus, a vertex appearing in multiple partitions implicitly encodes a inter-partition connection.

\stitle{Partition Distance.} 
To improve retrieval diversity and minimize redundancy, we aim to select partitions that are distant from one another in the original graph. A straightforward approach is to define a distance metric between every pair of partitions and select those that are far apart. However, computing such distances directly on the original graph via shortest paths is computationally prohibitive for large-scale graphs.\\
To address this, we introduce a lightweight graph abstraction that captures essential inter-partition connectivity while significantly reducing distance computational complexity. Specifically, we construct an auxiliary graph where each vertex represents a partition and edges indicate adjacency based on shared (replicated) vertex. This allows inter-partition distances to be approximated efficiently using standard graph traversal techniques.\\
To formally define adjacency between partitions, we first state a lemma from the node-replication model in Distributed-NE:
\begin{lemma}
If two partitions \(A\) and \(B\) are adjacent in the original graph \(G\), then \(V_A\cap V_B\neq\emptyset\).
\end{lemma}

\begin{proof}
Suppose \(A\) and \(B\) are adjacent in \(G\), so there exists an edge \((u,v)\in E\) with \(u\in A\), \(v\in B\). In edge partition methods, preserving this inter-partition edge requires replicating at least one of its endpoints into the other partition. Hence, either \(u \in B\) or \(v \in A\), which implies that \(V_A \cap V_B \neq \emptyset\).
\end{proof}
Using this result, we define an auxiliary graph to model partition-level relationships:
\begin{definition}[Partition Adjacency Graph]\label{PAG}
Let \(G=(V,E)\) be an undirected graph partitioned into \(\mathcal P=\{P_1,\dots,P_n\}\). The \emph{Partition Adjacency Graph (PAG)} is a graph \(G_H = (\mathcal{P}, E_H)\), where $E_H = \{(P_i, P_j) \mid P_i \cap P_j \neq \emptyset \}.$
\end{definition}

\begin{definition}[Distance in Partition Adjacency Graph]\label{hyper distance}
Let \(G_H=(\mathcal P, E_H)\) be the PAG as defined above. The distance \(d_H(P_i,P_j)\) between two partitions \(P_i\) and \(P_j\) is defined as the shortest path between \(P_i\) and \(P_j\) in \(G_H\).
\end{definition}

This abstraction offers two key advantages. First, it drastically simplifies distance computation by avoiding costly traversal on the full graph. Second, it enables the identification of structurally distant partitions, guiding the system to favor less-connected regions of the graph. These properties jointly improve both retrieval efficiency and the diversity of the selected results. At the end of this phase, pairwise structural distances between all partitions are available to support the subsequent selection process.

\noindent\textbf{Vertex Replication.} 
When a graph employs edge-partitioned, some edges inevitably span multiple partitions. To preserve connectivity, the endpoints of these inter-partition edges are replicated into the corresponding partitions. This mechanism, known as vertex replication, is widely adopted in existing partitioning methods \cite{49,48,2ps,50,bep}. 
In the following, we add a discussion on the vertex replication impact. We guarantee that replication does not affect the correctness of our results: (1) no results are missed, and (2) no duplicates remain. First, although results spanning multiple partitions could be lost, we address this by designing an inter-partition matching procedure (see Section 3.4), which ensures completeness. Second, replication may create duplicate matches, but such cases are rare. This mainly occurs in two scenarios: (i) a partition contains only replicated vertices without any original part of the result, which is highly unlikely under balanced partitioning; or (ii) two neighboring partitions independently discover the same partial match. In the latter case, we apply a deduplication strategy so that only one copy is retained. Furthermore, we demonstrate in Exp 2 (Table \ref{tab:ne-sheep}) that vertex replication has no impact on either effectiveness or efficiency, as verified by varying replication vertices and different replication ratios

\begin{figure}[htbp]
    \centering
\includegraphics[width=0.45\textwidth]{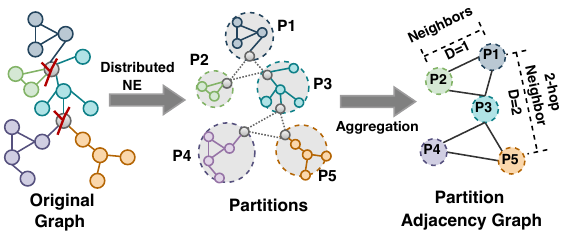}
    \caption{Illustration of Partitions Coarsen}
    \label{fig:pag}
\end{figure}

\begin{example}
    As illustrated in Figure~\ref{fig:pag}, the original graph is first partitioned using Distributed NE \cite{48} strategy, which introduces vertex cuts to divide the graph into multiple connected subgraphs. To maintain inter-partition connectivity, the vertex cuts are duplicated across the partitions they connect. Subsequently, for efficient distance estimation at the partition level, we coarsen the graph by abstracting each partition as a supernode. An edge is established between two supernodes if their corresponding partitions share at least one vertex replication, forming a Partition Adjacency Graph (PAG). This coarsened representation enables scalable computation of inter-partition distances, where the shortest path length between any two supernodes serves as an approximate distance between their respective subgraphs in the original graph.
\end{example}
\subsection{Distance-aware Partition Selection}
Our next objective is to select \( k \) structurally dispersed partitions from the partition set $P$. To promote distance diversity, we aim to select $k$ partitions from the set $P$ such that the selected partitions are well-separated in the graph. Specifically, we use the partition-level distance $d_H(P_i, P_j)$ defined in Definition~\ref{hyper distance} to measure the distance between two partitions. Our objective is to select a subset $S \subseteq \mathcal{P} \ \text{with}\ |S| = k$, in which the minimum pairwise distance among all partition pairs is maximized. 
This ensures that even the closest pair in S is still far apart, thus encouraging the overall dispersion of the selected regions. 
Although the pairwise distances between partitions can be efficiently computed using a preconstructed distance matrix, selecting the optimal subset of \( k \) partitions remains a combinatorial optimization problem. The search space consists of all possible \( k \)-sized subsets of \(\mathcal{P}\), totaling \({|\mathcal{P}| \choose k}\) combinations. As a result, exhaustive search becomes computationally infeasible on large-scale graphs.
For the solution, we employ a greedy dispersion-aware strategy to efficiently approximate the optimal solution. The procedure begins by randomly selecting an initial partition. In each step, we add the partition that has the maximum minimum distance to all partitions currently in the selected set. This process relies on a precomputed distance matrix and continues until \( k \) partitions are selected. The algorithm has a time complexity of \( O(kn) \), is simple to implement, and scales well to large graphs while producing well-dispersed partition sets.
\subsection{Hybrid Subgraph Matching}
After selecting the top-$k$ partitions, we perform fine-grained subgraph matching within them. Since partitioning phase may sever inter-partition structures, matches spanning multiple partitions may be missed. For this challenge, our algorithm use a two-phase matching strategy. In the first phase, a basic subgraph matching algorithm is applied independently within each partition to identify local matches. If no satisfactory result is found at this stage, the process advances to the second phase, in which the search is expanded across partition boundaries to recover connectivity and discover potential inter-partition matches.

The subgraph fine-matching procedure is detailed in Algorithm 2 and consists of two stages designed to ensure both completeness and result diversity.
In the first stage, each selected partition $G_i$ is evaluated independently and in parallel using a hybrid subgraph matching algorithm. 
We initialize candidate sets for each query vertex based on labels and local constraints, then refine them using bipartite graphs with semi-perfect matching checks ~\cite{gql} to prune invalid mappings. A cost model~\cite{gql} determines the join order, and LFTJ~\cite{lftj} is applied to compute synchronized intersections, enabling depth-first enumeration without large intermediates.

If no match is found within a partition, the algorithm enters the second stage to search results across partitions. 
In the second stage, BFS pattern trees are built, with each query vertex as root, to define the propagation order of constraints and guide inter-partition expansion. BFS cascading filtering is then applied from boundary candidate sets, retaining only inter-partition candidates that satisfy inexpensive checks (e.g., label consistency, degree bounds, adjacency summaries) while recording boundary information for stitching \cite{ceci}. The filtered frontiers are refined by DFS matching, where query vertices are incrementally mapped with forward checking and edge constraints. When traversing partitions, only essential states are preserved via stored boundary contexts to reduce overhead while ensuring correctness. Backtracking is invoked upon constraint violations to guarantee completeness.

\begin{algorithm}[hbp]
\small
\caption{\textsc{Hybrid Subgraph Matching (HySM)}}
\label{alg:2}
\SetAlgoLined
\SetAlgoNoEnd
\KwIn{Query graph $Q$, One of Selected partition $P_i$}
\KwOut{One matched subgraph in $P_i$ or \texttt{NoResult}}
\texttt{// Intra-partition matching} \\
Select and prune candidates with neighborhood information \cite{gql}; \\
Reduce search space using semi-perfect matching \cite{gql};\\
Optimize search order using a cost model \cite{gql}; \\
\If{LFTJ \cite{lftj} finds a match}{
    \Return{matched subgraph}
}

\texttt{// Inter-partition matching} \\
Generate BFS tree patterns from each vertex in $Q$\;
Identify cut-off candidates in $P_i$ for each vertex in $Q$\;
\ForEach{query vertex \textbf{in parallel}}{
    Perform BFS filtering to prune candidates\cite{ceci} \;
    Use DFS to find valid matchings from filtered candidates\;
    \If{valid match found}{
        \Return{matched subgraph}
    }
}
\Return{\texttt{NoResult}}
\end{algorithm}
\begin{theorem}
By maximizing the distance-based diversity objective function \(\mathcal{F}_{\text{dist}}(\mathcal{R})\), the framework simultaneously maximizes the coverage-based diversity \(\mathcal{F}_{\text{cov}}(\mathcal{R})\).
\end{theorem}
\begin{proof}
Let \(\mathcal{R} = \{c_1, c_2, \dots, c_k\}\) be the set of selected matches, where each \(c_i\) is chosen from a distinct partition \(P_i \in \mathcal{P}\). By the distance-based diversity criterion, the selected partitions are pairwise non-adjacent in the Partition Adjacency Graph (PAG). Hence, for all \(i \ne j\), we have \(P_i \cap P_j = \emptyset\), which implies \(V(c_i) \cap V(c_j) = \emptyset\).
It follows that the total coverage of the match set is equal to the sum of the sizes of the individual matches:
$\left| \bigcup_{i=1}^k V(c_i) \right| = \sum_{i=1}^k |V(c_i)|.$
Moreover, by construction, each \(c_i\) is selected to maximize \(|V(c_i)|\) within its corresponding partition \(P_i\). Since the vertex sets \(V(c_i)\) are disjoint and locally optimal within disjoint regions, no alternative selection of \(k\) disjoint subgraphs can yield a higher total coverage. Thus, the framework also maximizes \(\mathcal{F}_{\mathrm{cov}}(\mathcal{R})\).
\end{proof}
\begin{remark}
In our framework, the graph is partitioned with vertex replication to preserve inter-partition connectivity, so adjacent partitions share vertex cuts while non-adjacent ones do not (Lemma~3.1). Since PDD and PDD+ prioritize distance, typically non-adjacent partitions, the resulting matches contain no overlapping vertices. By Definition~2.5, coverage is defined as the union of matched vertex sets, which is maximized when matches are disjoint. Hence, ensuring distance-based diversity inherently guarantees maximum coverage, as further supported by our experimental results.
\end{remark}
\section{Optimizations} \label{sec:4}
The existing PDD framework faces two key limitations. First, selecting a fixed set of $k$ partitions without ensuring the presence of valid matches can lead to incomplete results and costly backtracking when matches are missing. Second, the commonly adopted greedy dispersion strategy lacks global optimality, as it relies on local decisions and is sensitive to initialization, often resulting in suboptimal and unstable partition selection.
To address these issues, we propose an optimized framework that improves both the effectiveness of selection and computational efficiency. In the following, we first present an overview of the optimized architecture and then provide detailed explanations and analysis of the two key improvements in the other two subsections.

\begin{algorithm}[t]
\small
\caption{The PDD Plus Framework}
\label{alg:optimized_framework}
\SetAlgoLined
\SetAlgoNoEnd
\KwIn{Data graph $G$, query graph $Q$, target size $k$}
\KwOut{Set $\mathcal{M}$ of $k$ diverse subgraph matches}
\tcp{Offline}
Execute \textit{Algorithm~\ref{alg:partition_matching}: Partition-based Graph Preprocessing};\\
Achieve $\{G_1,\dots,G_n\}$\& Distance matrix $D\in\mathbb{R}^{n\times n}$;\\
\tcp{Optimized Partition Selection}
\ForAll{partition $G_i$ \textbf{in parallel}}{
  $\mathbf{x}_i \gets \mathrm{ExtractFeatures}(Q, G_i)$\;
  $\hat y_i \gets \mathrm{Model.predict}(\mathbf{x}_i)$;
}
$\mathcal{P}_1 \gets \{\,G_i \mid \hat y_i = 1\}$\;
Construct graph $G_H=(\mathcal{P}_1, E_H, w_H)$ with
$E_H = \{(G_i,G_j)\mid G_i,G_j \in \mathcal{P}_1\},w_H(i,j)=D_{i,j}$
;
$\mathcal{M} \gets \mathsf{DekPS}(H, k)$ (Algorithm 4)\;
\Return{$\mathcal{M}$}
\end{algorithm}

As shown in Algorithm \ref{alg:optimized_framework}, the first stage follows Algorithm 1: Partition-based Graph Preprocessing phase. Partitions and their distance information are retained for downstream processing. The second stage introduces the use of representation learning from machine learning, where feature vectors are extracted for each subgraph. By embedding these vectors in an ordered space, we can efficiently determine the relationship between the query graph and each partitioned subgraph, estimating the probability of their existence. In the third stage, a Partition Distance Graph (PDG) is constructed by integrating both existence probabilities and inter-partition distances, enabling the discovery of a densest subgraph of size $k$. The detailed procedure is presented in Algorithm~\ref{Algor:interleaved}. 
Vertexes of achieved densest subgraph are the selected partitions. Finally, in the last stage, traditional algorithms are employed to perform parallelized subgraph matching within the top-$k$ ranked partitions. This approach optimizes the matching process by focusing computational resources on the most promising partitions.\\
\stitle{Time Complexity Analysis:}
Excluding offline preprocessing (graph partitioning and distance‐matrix computation), our online algorithm consists of two phases. First, feature extraction and model inference over all \(n\) partitions execute in parallel on \(n\) workers in \(O(1)\) wall‐clock time.  Second, we launch \(k\) threads for “select next” and fine‐grained subgraph matching: main thread scans \(m\) predicted candidates against the \(k\) already‐chosen partitions in \(O(mk)\) and the other $k$ threads performs matching in \(O(T_{\mathrm{match}})\) at the same time. Since main thread and all \(k\) threads run concurrently, this phase completes in \(O\bigl(\max\{m\,k,\;T_{\mathrm{match}}\}\bigr)\), which therefore also bounds the overall parallel time complexity.  

\begin{definition}[Approximation Ratio]
For distance-based diversity, we define the approximation ratio as
$\rho \;=\; \frac{D_{\text{alg}}}{D^*}$, where $D_{\text{alg}}$ is the distance-based diversity achieved by our algorithm, and $D^*$ is the optimal distance-based diversity.
\end{definition}

\begin{theorem}
Let $G_i$ and $G_j$ be two partitions in the Partition Adjacency Graph (PAG) such that the shortest path between them is $h = d_H(G_i, G_j)$.  Under the specific assumptions, the approximation ratio $\rho$ is bounded by:
$\rho\ge \frac{h-1}{h+1}$
\end{theorem}
\begin{proof}
Due to page limitations, the detailed proof is provided in the Appendix A (Theorem 2). 
\end{proof}
\subsection{Embedding-driven Partition Filtering}
In this phase, the central idea is to predict the most likely partitions of the query graph through the embedded domain relationships. To achieve this, we leverage the learned embeddings of both the query and the partitions, which encode structural and semantic features in a shared vector space. By comparing the query embedding with each partition embedding, we estimate the likelihood that a given partition contains potential matches. This similarity-based estimation allows for efficient pre-filtering of the search space, significantly reducing the number of partitions that need to be explored during the fine-grained matching phase. \\
\stitle{Initial Embedding.}
A key challenge in training arises from the scale mismatch between small query graphs and large partitions, which limits embedding comparability. Reducing partition size alleviates this but causes fragmentation and high overhead. To address this, we sample query-sized subgraphs from the data graph and pair them with queries as training instances, using higher sampling frequency on large graphs for coverage. We then employ GNNs to embed both queries and subgraphs into a shared space, capturing semantic and structural features for efficient comparison and alignment without costly structural matching.
\begin{figure} [t]
    \centering
\includegraphics[width=0.45\textwidth]{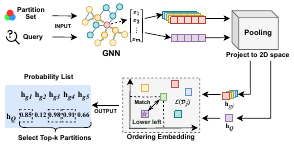}
    \caption{Neural Partition Filtering: Select $k$ likelihood Partitions(k=2).}
    \label{fig:enter-label_0}
\end{figure}

\stitle{Ordering Embedding.}
Graph data is inherently complex, and direct operations on the raw structure are often computationally expensive. Traditional Graph Neural Networks (GNNs) focus on messaging local neighborhoods information, but less effective at capturing global relationships—particularly subgraph inclusion. Against this backdrop, Order Embedding, proposed by Vendrov et al.\cite{60}, emerges as a suitable representation learning method which encodes partial order relations by enforcing component-wise inequalities between embedding vectors. \textbf{A prerequisite for applying order embedding is the explicit definition of partial order relations among the elements.}
\begin{property}[Partial Order]
Let \(S\) be a set. A binary relation \(\leq\) on \(S\) is called a \emph{partial order} if it satisfies the following three properties for all \(a, b, c \in S\): Reflexivity: \(a \leq a\); Antisymmetry: if \(a \leq b\) and \(b \leq a\), then \(a = b\); Transitivity: if \(a \leq b\) and \(b \leq c\), then \(a \leq c\).
\end{property}
\begin{theorem}
The subgraph matching problem naturally forms a partial order.
\end{theorem}
\begin{proof}
We verify the three conditions of partial order:
Reflexivity: Every graph is trivially a subgraph of itself; Antisymmetry: If \( G_1 \subseteq G_2 \) and \( G_2 \subseteq G_1 \), then \( G_1 = G_2 \) (i.e., the graphs are isomorphic). Transitivity: If \( G_1 \subseteq G_2 \) and \( G_2 \subseteq G_3 \), then \( G_1 \subseteq G_3 \). Therefore, the subgraph relation satisfies reflexivity, antisymmetry, and transitivity.
\end{proof}
Using this embedding, each pair (a subgraph and a query graph) is embedded into a continuous vector space, where the geometric relationships between the embeddings represent their partial order. Specifically, if the query graph is isomorphic to \( G_2 \), the embedding of the query graph is positioned geometrically ``below and to the left'' of the embedding of \( G_2 \) in the vector space. This spatial configuration is essential for capturing the properties of partial orders in the embedding space. 

\stitle{Training Process.}  
To learn embeddings that preserve partial order, we design a loss function that enforces component-wise inequalities between embedding vectors. For any graph pair $(G_1, G_2)$ where $G_1 \leq G_2$ (i.e., $G_1$ is a subgraph of $G_2$), their embeddings should satisfy:
$
\forall i,  E_{G_1}[i] \leq E_{G_2}[i].
$
Violations of this inequality are penalized through a loss function.\\
\underline{\textit{Positive Pairs.}}
For subgraph-supergraph pairs $(G_1, G_2)$, we apply a hinge-style loss: $L_{\text{pos}} = \sum_{i} \max(0, \; E_{G_1}[i] - E_{G_2}[i]).$\\
\underline{\textit{Negative Pairs.}}  
For unrelated pairs $(G_1, G_3)$ where $G_1 \nleq G_3$, we enforce a margin to prevent false ordering:$L_{\text{neg}} = \sum_{i} \max(0, \; \text{margin} + E_{G_3}[i] - E_{G_1}[i]).$\\
The overall training loss is defined as:
$
    \setlength\abovedisplayskip{0pt}
    \setlength\belowdisplayskip{0pt}
    L = \lambda_{\text{pos}} L_{\text{pos}} + \lambda_{\text{neg}} L_{\text{neg}},
$
where $\lambda_{\text{pos}}$ and $\lambda_{\text{neg}}$ balance the two components. Minimizing this loss across labeled graph pairs encourages the embedding space to reflect true subgraph relations. \\
\stitle{Inference and Filtering.}  
After training, the model is used to predict whether a query graph Q is a subgraph of a candidate graph G. Given their embeddings $E_Q$ and $E_G$, the model determines a positive subgraph relation (i.e., $Q \leq G$) if
$\forall i, E_Q[i] \leq E_G[i].$\\
At the beginning, the data graph is partitioned to identify regions likely to contain query matches.  Rather than evaluating each partition sequentially, we perform predictions in parallel across all partitions. This parallel inference strategy significantly reduces latency and accelerates candidate region identification during the online phase, with an accuracy of around 75\%.
Based on the prediction results, we treat partitions prediction labelled as 1 as those more likely to contain matching subgraphs. In the subsequent step, we select $k$ relatively dispersed partitions from this high-likelihood subset to ensure high-quality matching. This pre-selection strategy increases the likelihood that the final partitions contain valid results, thereby reducing the need for backtracking and improving overall efficiency.

\begin{remark}[Necessary Learning-based Strategy]
Although training a partition prediction model introduces additional computational overhead, it is a necessary and justifiable design choice. Once trained, the model can be reused across different queries and datasets with similar structural characteristics, making the cost both amortizable and transferable. 
\end{remark}
\subsection{Densest-based Partition Selection}
Based on the previously defined notion of diversity, our objective is to select $k$ partitions that are maximally distant from each other. Greedy-based approach to this problem often suffers from a lack of global optimality and strong sensitivity to the initial selection, resulting in suboptimal and unstable outcomes.
\begin{definition}[Densest Subgraph]
A subgraph $G_D$ of graph $G$ is called a densest subgraph if maximizes the density $\frac{|E_D|}{|V_D|}$. 
\end{definition}

\begin{definition}[Partition Distance Graph]\label{dif:dense}
Let \(G\) be partitioned into \(n\) disjoint blocks \(\mathcal{P}=\{P_1,\ldots,P_n\}\).
For each partition \(P_i\) is a vertex \(v_i\) in the \emph{partition distance graph} (PDG).
The PDG is then a weighted, undirected graph \(G_{\mathrm{P}} = (V, E, w)\), 
where \(V = \{v_1, v_2, \ldots, v_n\}\).
Two vertices \(v_i\) and \(v_j\) are connected by an edge in \(E\) if and only if their corresponding partitions 
\(P_i\) and \(P_j\) are non-adjacent in the original graph \(G\).  
For such a pair, the edge weight is
$w(v_i, v_j) = \mathrm{dist}_{\mathrm{PAG}}(P_i, P_j) - 1$,
where \(\mathrm{dist}_{\mathrm{PAG}}(P_i, P_j)\) denotes the shortest path between \(P_i\) and \(P_j\) in the partition adjacency graph (PAG).
\end{definition}

\begin{definition}[Degree in PDG]
The degree of a vertex \(v_i\) in the PDG is defined as the number of vertices to which it is connected,
i.e., the number of partitions \(P_j\) that are non-adjacent to its corresponding partition \(P_i\) in the original graph.
\end{definition}

To overcome these limitations, we reformulate the $k$ dispersed partition selection process as a densest subgraph discovery problem over PAG.
\begin{lemma}
Selecting a subset of \(k\) partitions that maximizes the minimum pairwise distance is equivalent to finding a \(k\)-size densest subgraph in the Partition Distance Graph (PDG) whose minimum edge weight is maximized.
\end{lemma}

\begin{proof}
We show that the original problem and the max–min edge-weight subgraph problem on the PDG share the same feasible space and objective.

\noindent\underline{\textit{Feasibility.}}
In the original problem, a valid solution requires selecting \(k\) partitions such that no two are adjacent in the data graph. In the PDG, vertices represent partitions, and edges exist only between non-adjacent pairs. Thus, any feasible subset of \(k\) partitions corresponds to a \(k\)-vertex subgraph in the PDG.

\noindent\underline{\textit{Objective.}}  
Let \(S \subseteq \mathcal{P}\) be a feasible selection. The original objective is to maximize
$
\mathcal{D}(S) = \min_{\substack{P_i, P_j \in S }} d(P_i, P_j).
$
In the PDG, edge weights are defined as \(w(P_i, P_j) = d(P_i, P_j)\), so the objective becomes
$
\mathcal{D}(S) = \min_{\substack{(P_i, P_j) \in E[S]}} w(P_i, P_j).
$
Hence, the original problem reduces to selecting a \(k\)-vertex subgraph of the PDG that maximizes the minimum edge weight.
\end{proof}
\begin{figure}[t]
    \centering  \includegraphics[width=0.45\textwidth]{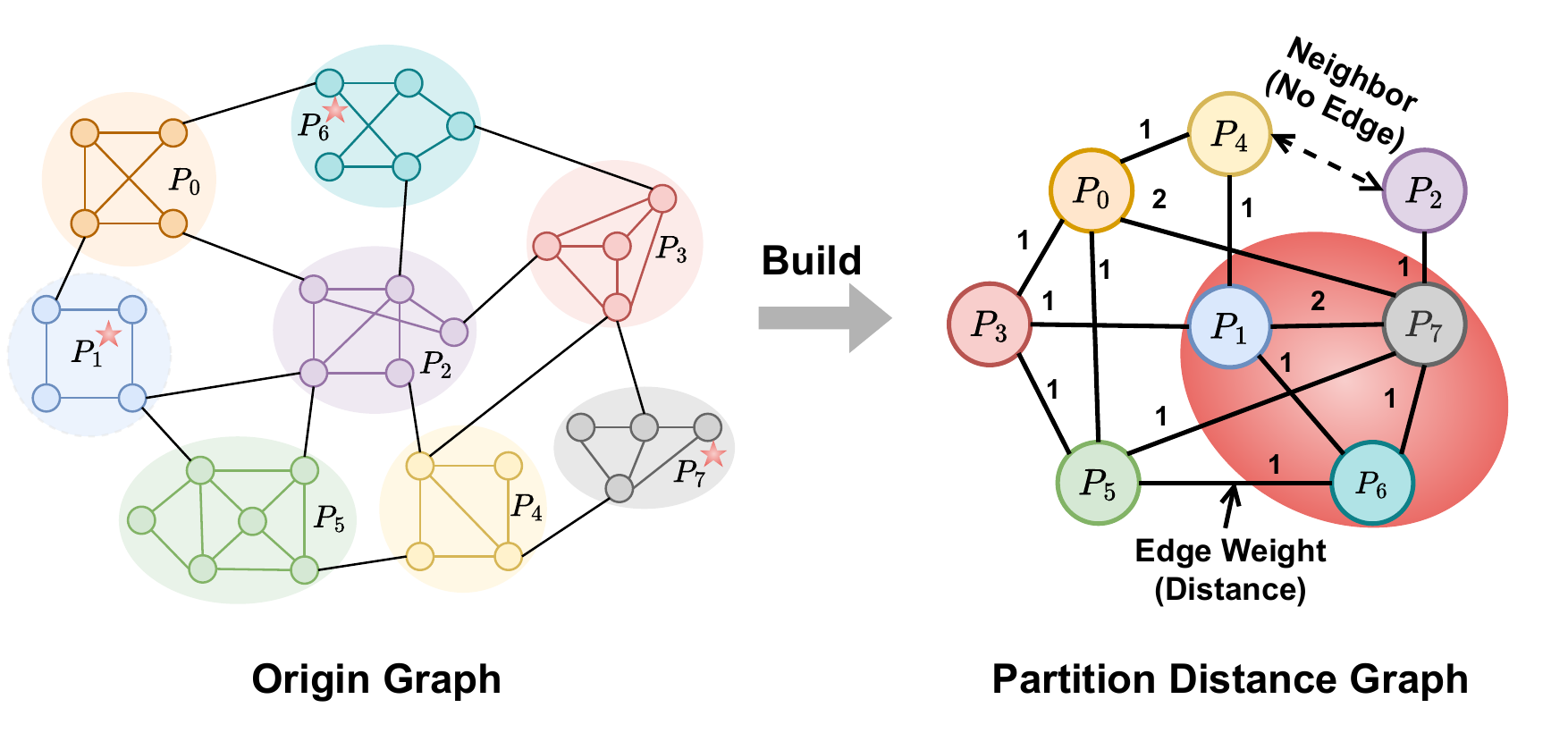}
    \caption{Dispersed $k$ Partitions Selection}
    \label{fig:densest}
\end{figure}

\begin{example}
    The overall process is illustrated in Figure~\ref{fig:densest}. In the graph preprocessing phase, the original data graph is partitioned, and both the adjacency relationships and pairwise distances between partitions are recorded. During the partition filtering stage, each partition is predicted as either likely or unlikely to contain any match of a query. Only those predicted as likely are retained as candidate partitions. Figure~\ref{fig:densest} visualizes only these high-likelihood candidates.
To enforce structural constraints and support efficient selection, we construct the Partition Distance Graph (PDG) as defined in Definition~\ref{dif:dense}.
%
We then search PDG for a $k$-size densest subgraph in which the minimum edge weight is maximized. In Figure \ref{fig:densest}, the highlighted region with a red background represents the selected subgraph. The partitions $p_1$, $p_7$, and $p_6$ are chosen as the $k=3$ most dispersed partitions based on this strategy. This transforms the diversity-driven selection problem into a densest subgraph discovery task: selecting a $k$-size densest subgraph from the PDG with the maximum total edge weight.
\end{example}

Compared to greedy methods, our approach captures global structure via a densest subgraph formulation over the PDG, enforces disjointness and dispersion through encoded constraints, and yields an interpretable $k$ size high-weight densest subgraph that reflects both coverage and distance diversity.

\stitle{Customized Densest subgraph discovery.} While classical densest subgraph algorithms aim to identify dense regions by exploring a variety of subgraphs with varying sizes, they are not directly applicable to our setting. We seek a single densest subgraph of fixed size $k$ that optimizes a specific objective—maximizing the each edge weight among selected vertices. Moreover, most existing algorithms are not designed to accommodate such task-specific goals.
To address these challenges, we design a task-specific heuristic that draws on established principles from densest subgraph discovery while accommodating our unique constraints.

\begin{algorithm}[t]
\small
\caption{\textsc{Densest-based \textit{k} Partition Selection (De$k$PS)}}
\label{Algor:interleaved}
\KwIn{PDG $G_P=(V,E,w)$, query graph $Q$, target size $k$, the candidate partition set $L$}
\KwOut{Set $\mathcal{M}$ of up to $k$ valid matches}
\BlankLine
\textbf{Initialization:}\\
$L \leftarrow \emptyset$; $\mathcal{M} \leftarrow \emptyset$; \\
$p^{\ast} \leftarrow \arg\max_{v\in V}\deg(v)$;\\
$L \leftarrow L \cup \{p^{\ast}\}$;\\
$\mathcal{M}_{p^{\ast}} \leftarrow \mathsf{HySM}(Q, p^{\ast})$ (Algorithm 2);\\
\If{$\mathcal{M}_{p^{\ast}} \neq \emptyset$}{
    $\mathcal{M} \leftarrow \mathcal{M} \cup \mathcal{M}_{p^{\ast}}$;
}
$N \leftarrow \bigcap_{v \in L} Neighbor(v) \setminus L$;
\BlankLine
\While{$|\mathcal{M}| < k$ \textbf{and} $N \neq \emptyset$}{
  $p^{\ast} \leftarrow \arg\max_{c \in N} \left( \max_{l \in L} w(c, l) \right)$;\\
  \If{multiple candidates}{
    choose $p^{\ast}$ with highest degree among candidates;
  }
  $L \leftarrow L \cup \{p^{\ast}\}$ ;\\
  $\mathcal{M}_{p^{\ast}} \leftarrow \mathsf{HySM}(Q, p^{\ast})$ \textbf{in parallel}
 (Algorithm 2);\\
  \If{$\mathcal{M}_{p^{\ast}} \neq \emptyset$}{
      $\mathcal{M} \leftarrow \mathcal{M} \cup \mathcal{M}_{p^{\ast}}$;
  }
  $N \leftarrow \bigcap_{v \in L} Neighbor(v) \setminus L$;
}
\Return{$\mathcal{M}$}
\end{algorithm}

As shown in Algorithm \ref{Algor:interleaved}, \textsc{De$k$PS} heuristically selects $k$ partitions from the PDG to form a high-weight, structurally disjoint subgraph. The strategy is inspired by densest subgraph discovery, aiming to incrementally construct a size-$k$ subgraph with maximal edge weight under non-adjacency constraints. The algorithm starts from the partition with the highest degree and adds it to the candidate set $L$. At each iteration, it identifies candidate partitions that are non-adjacent to all members of $L$ and selects the one that contributes the highest edge weight to $L$, thus approximating a dense, dispersed subgraph. Ties are resolved by degree. Each selected partition undergoes subgraph matching via the hybrid matcher $\mathsf{HySM}$ (Algorithm 2), which is executed in parallel across partitions. The process continues until $k$ valid matches are found or no further candidates remain. Through a peeling-style expansion guided by edge weights, \textsc{De$k$PS} balances the goals of structural density and global dispersion in a unified selection framework.

\section{Experiments} \label{sec:5}
In this section, we introduce the experimental setup, including the datasets, queries, and baselines used in the experiments. We then present the preprocessing cost and the performance of our algorithm in terms of efficiency, effectiveness, and scalability.

\stitle{Environment.}
Experiments are conducted on a machine equipped with an Intel(R) Xeon(R) Gold 5218R CPU @ 2.10GHz, 256GB main memory, and a NVIDIA Tesla T4 GPU. We use PyTorch and C++ for the algorithm implementation.

\stitle{Datasets.} 
We evaluate our approach using real-world and synthetic graph datasets and compare its performance with baseline methods. Detailed information on all datasets is provided in Table \ref{tab:datasets}.
\underline{\textit{Real-World Data Graph.}} We use 10 real-world graph datasets from prior studies \cite{72}. 
\noindent\underline{\textit{Synthetic Data Graphs.}}
Existing real-world datasets are relatively small to evaluate scalability, we follow prior work~\cite{2, ldbc2, mgm, sus, pcc, daf} and use LDBC \cite{ldbc} to generate synthetic datasets. The synthetic dataset, denoted as DG$x$ ($x$ is the scale factor), includes integer node labels from 0 to 10, supporting reproducible and scalable evaluation.

\begin{table}[t]
  \centering
  \caption{Statistics of Datasets (\underline{Dataset Generated Labels}).}
  \vspace{0.5em} 
  \label{tab:datasets}
  \fontsize{7}{9}\selectfont
  \begin{tabular}{
    >{\centering\arraybackslash}p{1.2cm}
    >{\centering\arraybackslash}p{0.5cm}
    >{\centering\arraybackslash}p{1.2cm}
    >{\centering\arraybackslash}p{1.2cm}
    >{\centering\arraybackslash}p{0.4cm}
    >{\centering\arraybackslash}p{0.6cm}
    >{\centering\arraybackslash}p{0.8cm}}
    \hline
    \multicolumn{1}{c}{\textbf{Datasets}} & 
    \multicolumn{1}{c}{\textbf{Name}} & 
    \multicolumn{1}{c}{\textbf{$|V|$}} &  
    \multicolumn{1}{c}{\textbf{$|E|$}} & 
    \multicolumn{1}{c}{\textbf{$|\Sigma|$}} & 
    \multicolumn{1}{c}{\textbf{Degree}} & 
    \multicolumn{1}{c}{\textbf{Type}} \\
    \hline
    Yeast & \textit{ye} & 3,112 & 12,519 & 71 & 8.0 & Protein \\
    Human & \textit{hu} & 4,674 & 86,282 & 44 & 36.9 & Protein \\
    HPRD & \textit{hp} & 9,460 & 34,998 & 307 & 7.4 & Protein \\
    \underline{FreeBase15k} & \textit{fb} & 14,951 & 260,184 & 1 & 42.7 & Lexical \\
    \underline{Wordnet18} & \textit{wm} & 40,943 & 75,770 & 4 & 3.7 & Lexical \\
    WordNet & \textit{wn} & 76,853 & 120,399 & 5 & 3.1 & Lexical \\
    DBLP & \textit{db} & 317,080 & 1,049,866 & 15 & 4.9 & Social \\
    \underline{DBpedia} & \textit{dp} & 343,794 & 1,371,562 & 1233 & 8.0 & Wiki \\
    Youtube & \textit{yt} & 1,134,890 & 2,987,624 & 24 & 5.2 & Social \\
    US Patents & \textit{up} & 3,774,768 & 16,518,947 & 25 & 8.8 & Citation \\
    \underline{DG 10} & \textit{d10} & 29,987,834 & 176,481,399 & 11 & 11.6 & Synthetic \\
    \underline{DG 60} & \textit{d60} & 187,108,072 & 1,246,659,788 & 11 & 13.2 & Synthetic \\
    \hline
  \end{tabular}
\end{table}

\begin{figure}[htbp] 
    \centering  
    \includegraphics[width=0.5\textwidth]{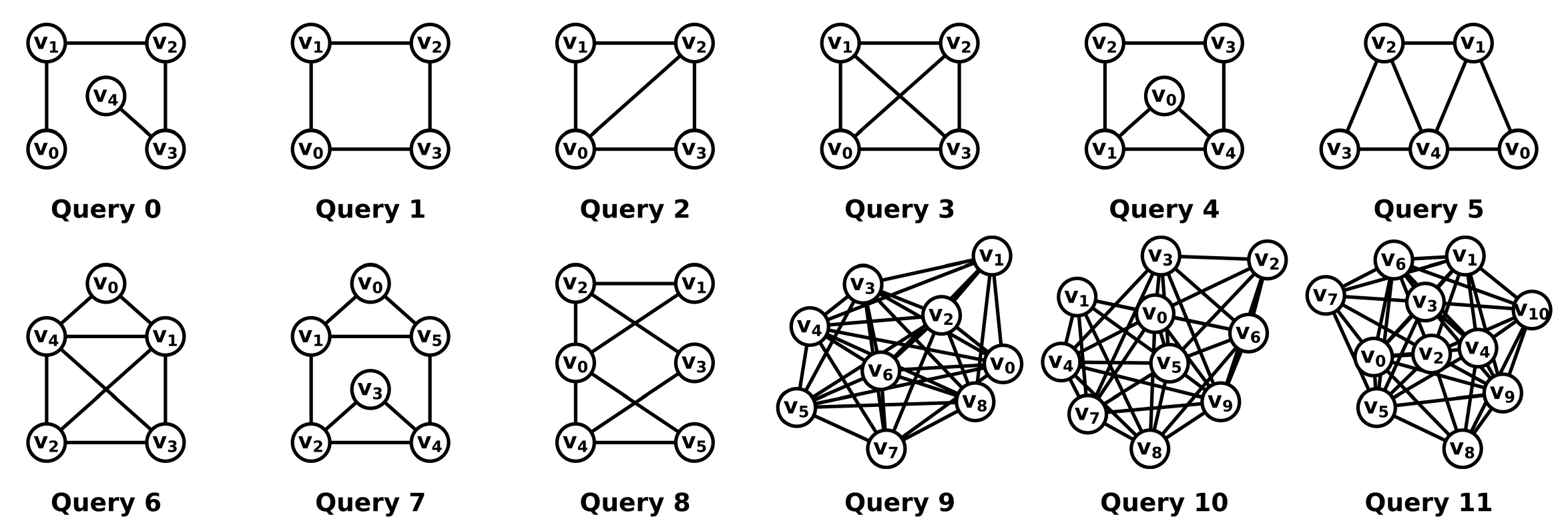} 
    \caption{Examples of different kinds of queries.} 
    \label{fig:query}
\end{figure}

\stitle{Query.} Traditional random-walk methods often yield overly simple query graphs. To ensure controlled and diverse evaluation, we categorize queries by topological complexity: \textit{simple} (paths, cycles), \textit{common} (combinations with shared vertices/edges), and \textit{complex} (dense patterns such as quasi-cliques). Queries are generated using random walks, frequent subgraph mining, and densest subgraph discovery. Each set contains 1,000 queries with fewer than 20 vertices. Representative patterns are shown in Figure~\ref{fig:query}.\\
\stitle{Baselines.} We evaluate 11 algorithms: our proposed methods \textit{PDD} and \textit{PDD+}, along with 9 baselines. 
DSQL \cite{73} is a diversified top-$k$ subgraph matching algorithm, while PTAB \cite{ptab} is a top-$k$ subgraph matching algorithm. GuP \cite{gup}, CFL \cite{cfl}, GQL \cite{gql}, DAF \cite{daf}, GQLfs \cite{gqlfs}, RapidMatch (RM) \cite{rm}, and VEQ \cite{veq} are subgraph matching algorithms, among which GuP and DAF support parallel execution.

\subsection{Initial Partition Evaluation}
To give a comprehensive view of the offline phase, we present detailed partitioning information such as partition costs, and conduct a robustness analysis to examine the impact of different partitioning schemes on efficiency and result quality.\\
\begin{table*}[htbp]
\vspace{-16pt}
\centering
\caption{Preprocessing cost on different datasets}
\label{tab:1}
\fontsize{7}{9}\selectfont
\begin{tabular}{lccccccccccccc}
\toprule
Dataset & Yeast & HPRD & Human & FB-15K & WordNet18 & WordNet & DBLP & DBpedia & YouTube & US-Patents & YAGO & DG10 & DG60 \\
\midrule
Graph Size(MB) & 0.11 & 0.31 & 0.74 & 2.47 & 0.83 & 1.32 & 13.29 & 17.89 & 36.93 & 249.54 & 2590.72 & 3001.92 & 23490.56 \\
Partition Time(s) & 0.06 & 0.08 & 0.10 & 0.19 & 0.12 & 0.18 & 0.59 & 1.05 & 2.52 & 19.32 & 167.43 & 148.86 & 1078.82 \\
Partition Memory(MB) & 8.11 & 8.52 & 8.76 & 14.25 & 9.57 & 12.59 & 62.95 & 77.25 & 174.57 & 883.00 & 11622.00 & 9044.00 & 63693.00 \\
\bottomrule
\end{tabular}
\end{table*}
\begin{figure*}[!t]
  \centering
  \begin{minipage}[t]{0.32\textwidth}
    \centering
    \includegraphics[width=\linewidth]{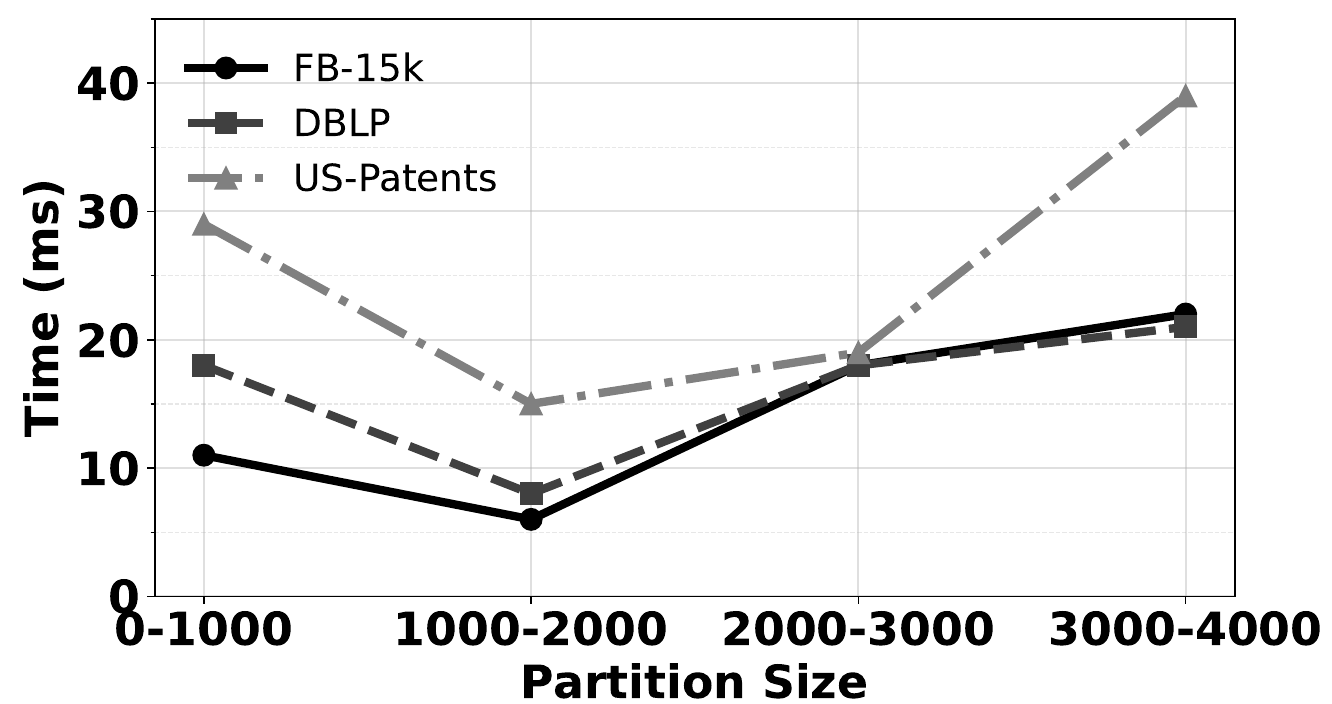}
    \caption{Partition Size Evaluation.}
    \label{fig:aaa1}
  \end{minipage}
  \hfill
  \begin{minipage}[t]{0.32\textwidth}
    \centering
    \includegraphics[width=\linewidth]{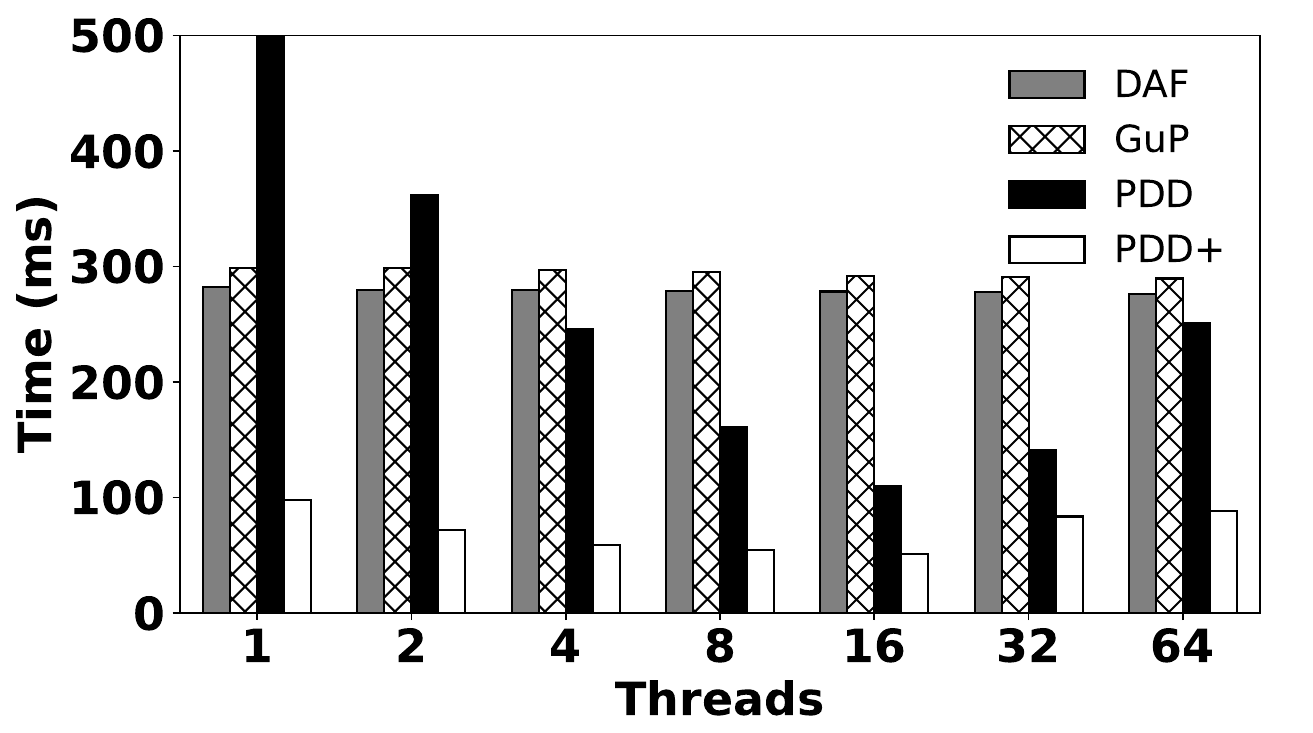}
    \caption{Parallel Comparison (YouTube).}
    \label{fig:aaa2}
  \end{minipage}
  \hfill
  \begin{minipage}[t]{0.32\textwidth}
    \centering
    \includegraphics[width=\linewidth]{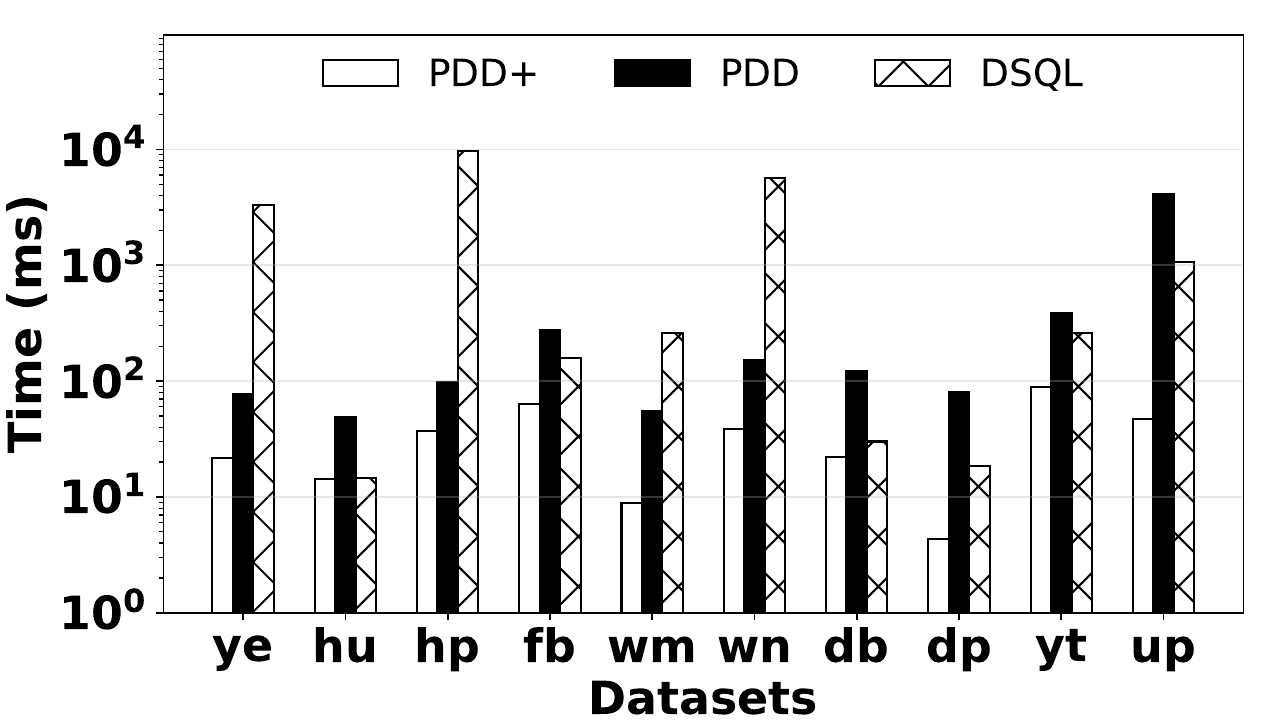}
    \caption{Sequential Comparison.}
    \label{fig:aaa3}
  \end{minipage}
\end{figure*}
\noindent\textbf{Exp1 - Preprocessing Cost.} 
We adopt a default partition size in the range of  1000–2000 vertices, under which we report the corresponding preprocessing time and memory costs across all datasets in the Table \ref{tab:1}. On most datasets, the partitioning time is only a few seconds with memory costs in the MB scale. For million-scale graphs, the overhead remains modest. Although the partitioning time becomes relatively high on billion-scale datasets, it is a one-time offline process and therefore acceptable. The choice of partition sizes in the range of 1000–2000 vertices is motivated by the favourable trade-off it offers between efficiency and result quality, and the experimental results are shown in Figure \ref{fig:aaa1}. Smaller partitions lead to excessive fragmentation and costly inter-partition searches, whereas larger ones approximate the entire graph and thereby reduce parallelism.\\
\begin{table}[htbp]
\fontsize{7}{9}\selectfont
\centering
\caption{Comparison of different partition schemes.}
\label{tab:ne-sheep}
\setlength{\tabcolsep}{3pt} 
\begin{tabular}{l|ccc|ccc|ccc}
\toprule
\multirow{2}{*}{\textbf{Dataset}} & 
\multicolumn{3}{c|}{\textbf{Replication Ratio}} & 
\multicolumn{3}{c|}{\textbf{Execution Time(ms)}} & 
\multicolumn{3}{c}{\textbf{Result Distance }} \\
\cline{2-10}
& NE & SHEEP & Leiden & NE & SHEEP & Leiden & NE & SHEEP & Leiden \\
\midrule
HPRD    & 2.57 & 2.12 & 1.00 & 1.65 & 1.48 & 5.80 & 2.46 & 2.56 & 2.06 \\
DBLP    & 1.43 & 1.88 & 1.00 & 1.64 & 1.81 & 21.01 & 5.95 & 5.93 & 5.61 \\
YouTube & 1.77 & 2.21 & 1.00 & 25.77 & 25.63 & 31.25 & 4.58 & 4.60 & 3.71 \\
\bottomrule
\end{tabular}
\end{table}
\noindent\textbf{Exp 2 - Robustness under Partitioning Schemes.} \label{exp2}
We evaluated different partitioning methods to analyze the robustness of our algorithm in terms of execution time and result quality. Specifically, the original graph was partitioned using two edge partitioning strategies (SHEEP \cite{49} and Distributed-NE \cite{48}) as well as a community detection method (Leiden) \cite{leiden}, and the resulting partitions were then used as inputs to our framework.
As shown in Table~\ref{tab:ne-sheep}, although the edge-based methods yield different replication ratios, our framework produces almost identical outcomes in both efficiency and effectiveness. The maximum runtime difference between them is only 0.17 ms, while the distance diversity of the results differs by at most 0.12, indicating that our algorithm is robust and largely unaffected by the choice of edge-based partitioning scheme. 
In contrast, the community detection method performs worse in both matching time and distance diversity compared with edge-based approaches. For example, on the \textit{DBLP} dataset, it is about 20 times slower. This is because, unlike edge-based partitioning, community detection does not replicate endpoints: each vertex belongs to exactly one community (partition). As a result, matches that involve boundary vertices may be missed, leading to excessive inter-partition searches and degraded efficiency.

\subsection{Efficiency Evaluation}
Using the default k=30, we compared the execution times of our algorithm with baselines on real and synthetic datasets, evaluating both parallel and single-thread efficiency.

\noindent\textbf{Exp3 - Overall Efficiency Comparison (Real-World Data).} 
\begin{figure*}[htbp]
    \centering
    \includegraphics[width=0.98\linewidth]{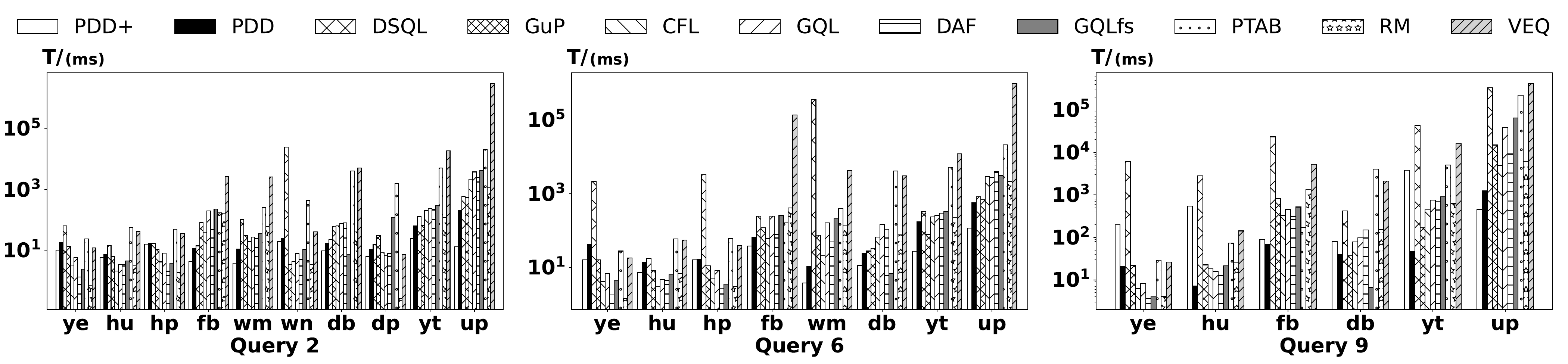}
    \caption{Runtime of Different Query on Different Real-World Datasets (\textit{k} = 30)}
    \label{fig:3}
\end{figure*}
As shown in Figure~\ref{fig:3}, we evaluate PDD+ and baselines across all queries and real-world datasets. For clarity, we present results on three representative queries: one simple query, one medium-complexity query, and one near-clique query. The remaining experimental results are provided in the Appendix B.
%
For simple queries (Query 2), PDD and PDD+ generally achieve lower runtimes across most datasets, with clear advantages on large graphs such as \textit{yt} and \textit{up}. As dataset size grows, baselines (e.g., CFL, GuP, PTAB, VEQ) often escalate to $10^5$ms, while PDD and PDD+ remain efficient. PDD+ also consistently outperforms DSQL(another diversified top-k subgraph matching algorithm), reducing latency by about 70\% and achieving up to three orders of magnitude speed-up on sparse graphs (e.g., \textit{wm}, \textit{wn}), demonstrating robustness across data distributions. 

On smaller datasets (\textit{ye}–\textit{hp}), some classical algorithms (e.g., CFL, GQL, DAF, RM) may run faster since the partition graph is similar in size to the original graph, making their processing time comparable to our intra-partition matching. However, these methods provide very poor diversity (see Section 5.2), whereas PDD+ again surpasses them on larger datasets.
For moderately complex queries (Query 6), PDD+ achieves over an order-of-magnitude speed-up compared to DSQL, and up to four orders on sparse datasets such as \textit{wm}, highlighting its robustness as query complexity increases. While classical methods may occasionally run faster on small graphs, PDD+ consistently outperforms them on larger datasets. For near-clique queries (Query 9), PDD+ still achieves two to three orders of magnitude speed-up over DSQL. Compared with subgraph enumeration, PDD+ is not always the fastest on small graphs, but consistently shows clear advantages on medium and large datasets.

\noindent\textbf{Exp4 - Overall Efficiency Comparison (Synthetic Data).} \\
\begin{figure}[htbp]
    \centering
    \includegraphics[width=0.95\linewidth]{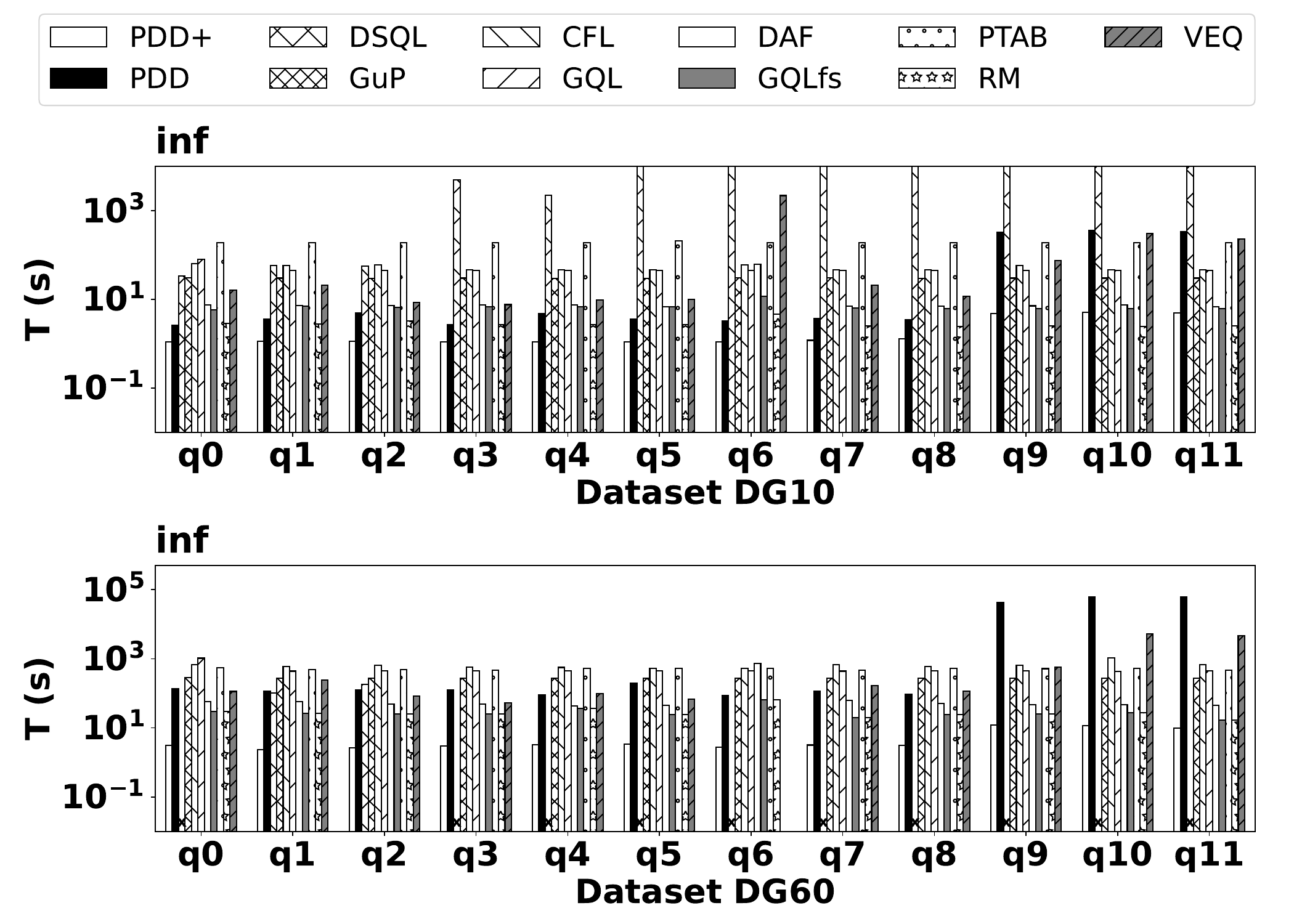}
    \caption{Runtime of All Queries on Syn-Datasets (\textit{k} = 30)}
    \label{fig:syn-dataset}
\end{figure}
We also evaluate all queries on two synthetic datasets, \textit{DG10} and \textit{DG60}. Due to the extreme scale of these datasets, DSQL sometimes doesn't run properly due to its extensive record of temporary data for pruning during the process.
As shown in Figure~\ref{fig:syn-dataset}, PDD+ and PDD are generally among the fastest methods. On the large-scale dataset DG60, PDD+ is typically one to two orders of magnitude faster than most baselines, and in clique queries (q9–q11) it can maintain advantages exceeding three orders of magnitude.
DSQL is clearly slower than PDD+. On DG10 the gap is about one order of magnitude, while on DG60 with more complex queries (q6–q11), the difference further enlarges to two or even three orders of magnitude.
Subgragh enumeration methods (CFL, GuP, GQL, DAF, etc.) are occasionally comparable to, or slightly faster than, PDD+ on small datasets and simple queries (q0–q2). However, as query complexity grows, their runtime increases sharply, and in clique queries (q9–q11) they are often two to three orders of magnitude slower than PDD+.
DSQL, PTAB, VEQ  remain the slowest across all baselines, being consistently two to three orders of magnitude slower than PDD+ on DG10, with the gap widening further (up to three to four orders) on DG60.

\noindent\textbf{Exp5 - Parallel Comparison.} The experiment evaluates the parallel performance of our method against other parallelizable subgraph matching algorithms under varying thread counts using Query 1. Since existing diversified top-$k$ subgraph matching algorithms do not support parallel execution, 
we compare our method against parallelizable subgraph matching algorithms including DAF and GuP. In our framework, the partition-based design enables parallelism throughout the entire process, 
whereas DAF and GuP apply parallelism only during the backtracking stage. As shown in Figure \ref{fig:aaa2}, PDD+ consistently outperforms the baselines, being about three times faster under fewer threads and reaching up to 6× speed-up as the thread count increases. For more details, our optimized algorithm achieves sub-100 ms runtime across all thread settings, whereas the baselines remain around 300 ms. PDD is slower than the baselines in the fewer-thread case, since it explores many partitions without results and incurs additional costs for inter-partition tree reconstruction. However, with more threads, these costs are effectively amortized through parallelism, allowing PDD to gradually surpass the baselines.\\
\noindent\textbf{Exp6 - Single-Thread Comparison.} The experiment compares the single-thread performance of our algorithm with DSQL, a diversified top-$k$ subgraph matching algorithm with the same problem definition, across various real-world datasets using Query~1 to evaluate efficiency. As shown in Figure \ref{fig:aaa3}, the sequential version of PDD+ consistently achieves the lowest runtime, often 1–3 orders of magnitude faster than DSQL. On datasets with fewer results (e.g., \textit{hp}, \textit{wn}, \textit{up}), sequential PDD incurs higher costs by frequently exploring partitions without feasible matches, whereas PDD+ incorporates a prediction stage that avoids such overhead by filtering out infeasible partitions in advance.
\subsection{Effectiveness Evaluation}
\begin{figure*}[htbp]
\vspace{-20pt}
\centering
\includegraphics[width=1.0\linewidth]{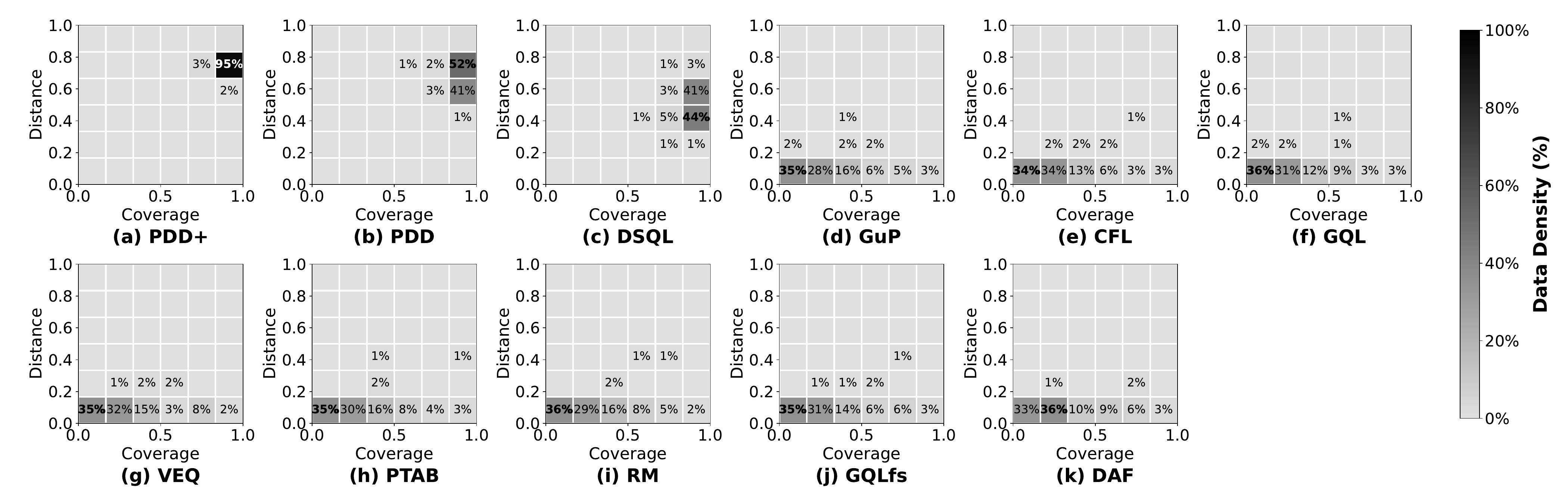}
    \caption{Coverage vs. Distance Metrics Across Valid Datasets and Queries }
    \label{fig:heat}
\end{figure*}
\noindent\textbf{Exp7 - Results Quality Evaluation.}
We analyze the \textit{coverage} – \textit{distance} distribution patterns for PDD+ and baselines in Figure~\ref{fig:heat}. For \textit{coverage} (Definition \ref{cover dist}), it is normalized by its exact optimal value—the highest \textit{coverage} over all combinations of matches. For \textit{distance} (Definition \ref{def:distance}), we conduct a greedy algorithm with backtracking on all matches to calculate the approximate best distance to normalize \textit{distance} metric.
As depicted in Figure~\ref{fig:heat}, about 95\% of the PDD+ matches fall in the upper right region, indicating simultaneously high normalized \textit{coverage} and high normalized \textit{distance}, and thus superior diversity. In contrast, PDD achieves a comparable quality in only 51\% of cases, with the rest exhibiting low normalized \textit{distance}. DSQL performs even worse: despite generally high normalized \textit{coverage}, its matches have low \textit{distance} because it considers only vertex coverage. Other subgraph enumeration algorithms also lack diversity guarantees. In about 35\% of the cases—the largest proportion—their top-$k$ matches predominantly occupy the lower-left region, indicating both low coverage and distance diversity.

\begin{figure}[htbp]
  \centering
  \includegraphics[width=0.5\textwidth]{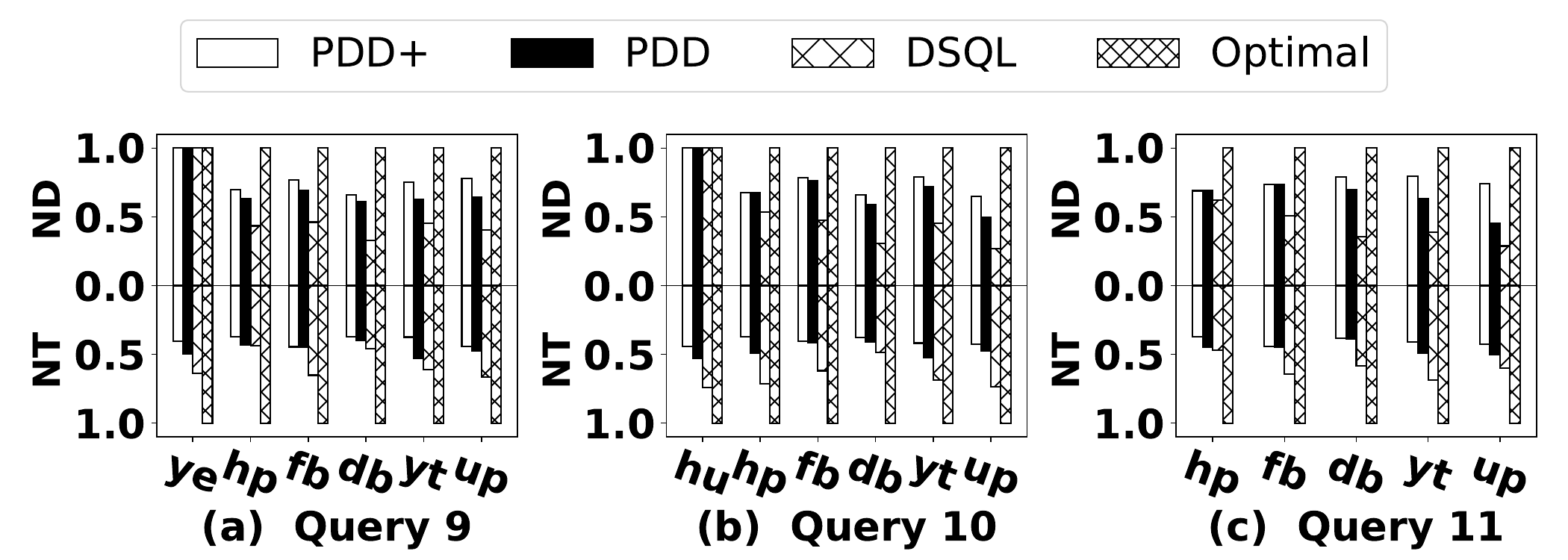}
  \caption{Distance–Time Trade‐Off on Complex Queries.}
  \label{fig:distance_time_tradeoff}
\end{figure}

\noindent\textbf{Exp8 - Distance-Time Trade-Off Evaluation.}
In Figure~\ref{fig:distance_time_tradeoff}, we report the trade-off between diversity (distance) and efficiency (time), comparing PDD+ with PDD, DSQL, and the optimal distance value. For a fair baseline, we adapt existing subgraph matching methods to account for topological dispersion and treat them as the “Optimal” algorithm: first enumerating all matches with a subgraph isomorphism algorithm, then computing pairwise distances and selecting the most dispersed $k$ results. Both distance and time are normalized for comparability.
To ensure fair comparison, both distance and time are normalized. Specifically, the normalized distance is defined as $ND = dis/ dis_E$, where $dis_E$ is the best achievable distance. For time, we use $NT = \log_{10}(time) / \log_{10}(time_E)$, where $time_E$ is the elapsed time to compute the best distance. This normalization highlights the trade-off between solution diversity and efficiency across different algorithms.
In Figure \ref{fig:distance_time_tradeoff}, we observe that PDD+ always uses less elapsed time to obtain a higher distance compared to PDD, DSQL, and the best distance computation. 
Overall, PDD+ achieves a favourable trade-off, reaching 65--80\% of the optimal distance while being up to six orders of magnitude faster than exact computation.
Compared to PDD, PDD+ is about 7$\times$ faster on average in runtime and achieves a 14.7\% improvement on average in distance, demonstrating the effectiveness of our optimizations.  
Compared to DSQL, PDD+ is approximately 1725$\times$ faster in runtime and achieves at least a 76.8\% improvement in distance, highlighting its superiority in both efficiency and diversity.
For small-result cases (e.g., Query~9 on \textit{ye} and Query~10 on \textit{hu}), where fewer than 30 matches exist, both PDD+ and DSQL are able to enumerate all matches, thus reaching the maximum distance.

\subsection{Scalability Evaluation}
Since our problem involves diversified top‑k matches, we assess scalability by varying k and by scaling the data at k=30. With space constraints, we report results from three datasets of different sizes and three representative query types. We compare our method to DSQL and representative classical baselines (CFL, RapidMatch), omitting other baselines with similar trends; additional results are detailed in the Appendix B.

\begin{figure}[htbp] 
    \centering 
    \includegraphics[width=0.475\textwidth]{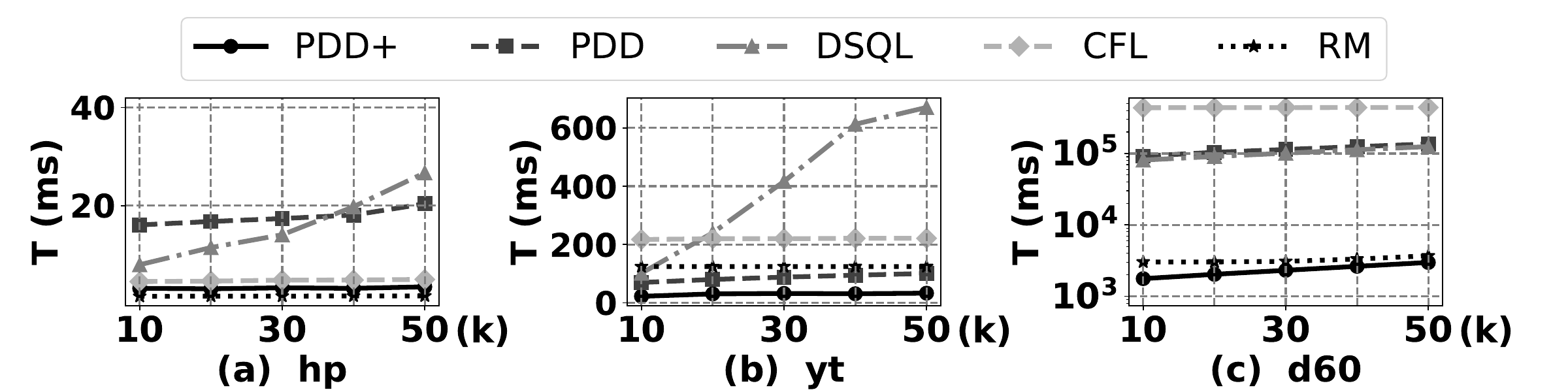} 
    \caption{Runtime Variation with \textit{k} for Query q1.} 
    \label{fig:time-k}
\end{figure}

\noindent\textbf{Exp9 - Time-\textit{k} Evaluation.} 
We vary the value of $k$ from 10 to 50, with a small real-world dataset, a large real-world dataset, and a synthetic one to show changes in runtime.
As shown in Figure \ref{fig:time-k}, PDD+ consistently achieves the fastest runtime across different real-world datasets and synthetic datasets, remaining nearly unaffected as $k$ increases. PDD also shows stable runtime trends, but it is consistently slower than PDD+. RM demonstrates similarly fast performance and its runtime is almost insensitive to the value of $k$. In contrast, CFL exhibits relatively poor performance on several datasets, yet its runtime grows smoothly with $k$. DSQL, however, shows the most significant increase, with runtime escalating rapidly on most datasets as $k$ becomes larger.

\begin{figure}
    \centering
    \includegraphics[width=1.0\linewidth]{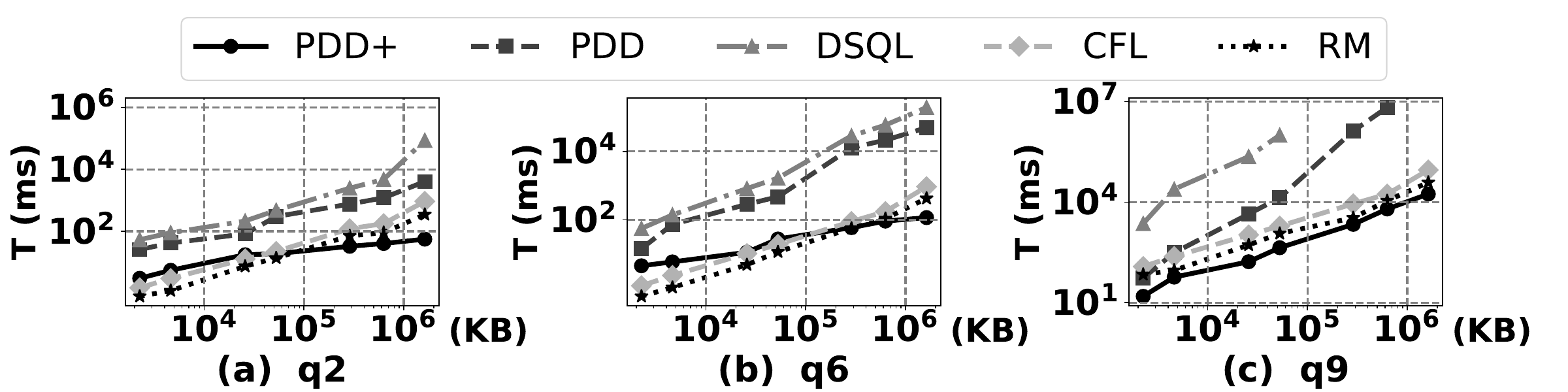}
    \caption{Runtime Variation with Data Size (k = 30).}
    \label{fig:datasize-k}
\end{figure}

\stitle{Exp10 - Datasize-\textit{k} Evaluation.} 
To analyze the scalability of PDD+ while reducing structural bias, we extract seven sub-datasets of varying sizes from DG10  using a simple query, a medium-complexity query, and a near-clique query to demonstrate trends of runtime. Overall, as shown in Figure \ref{fig:datasize-k}, PDD+ achieves the best performance in most cases, with the lowest runtime and smooth growth. PDD also grows relatively steadily, whereas DSQL increases sharply with $k$, becoming orders of magnitude slower on large datasets. Subgraph matching baselines remain relatively stable. CFL and RM achieve low runtimes on small datasets. As the graph size increases, performance gaps between algorithms become more pronounced, with the superiority of PDD+ most evident under logarithmic scales.
In summary, these experimental results demonstrate that the performance of PDD+ remains stable when faced with varying \textit{k} values and data scales, and it can process large-scale data without encountering significant performance bottlenecks.

\section{Related Work} \label{sec:6}
\stitle{Graph Partitioning.}
Graph partitioning enables scalable graph computation by decomposing large graphs into smaller subgraphs. Approaches fall into three categories: vertex partitioning \cite{52, 46, 47}, which minimizes edge cuts but struggles with high-degree nodes; edge partitioning \cite{48, 49, 50, 51}, which replicates vertices to improve load balance; and hybrid partitioning \cite{53, 54, 55}, which combines both to balance computation and communication.

\stitle{Densest Subgraph.}
The densest subgraph problem is widely studied due to its applications in biology, summarization, and anomaly detection. Exact algorithms \cite{68, 71} ensure optimality, while approximate methods \cite{69, 70, 71} use greedy strategies to scale to large graphs. Variants extend to constraint-aware, objective-tuned, streaming, and k-clique settings \cite{71}.

\noindent\textbf{Subgraph Matching:} Subgraph matching finds isomorphic embeddings of a query graph in a data graph. Classical methods adopt a filtering-ordering-enumeration paradigm \cite{44, 72}. These approaches employ optimized filtering strategies to shrink candidate sets and carefully designed matching orders to further reduce the search space. For enumeration, some algorithms \cite{gup, gql, vf2+, ceci, cfl, 7907163, daf, gqlfs} enumerate subgraphs using backtracking strategies, and some of these can be parallelized, such as DAF \cite{daf} and GuP \cite{gup}. Other algorithms \cite{dualsim, mapreduce, sde, qo2, 9417730, huge, ptab} enumerate results using joining strategies. For example, RapidMatch \cite{rm} leverages nucleus decomposition to improve the efficiency of the join phase. There are also learning-based methods to achieve subgraph matching, including both approximate \cite{neuralmatch,67} and exact \cite{61,64,gnndual} approaches.

\noindent\textbf{Result Diversification:} Diversified result selection balances relevance and diversity to avoid redundant output, and it is widely applied in information retrieval \cite{ir1, ir2, ir3, ir4, ir5}, recommendation systems \cite{rd1, rd2, rd3, rd4, rd5}, database systems \cite{74, 76} and subgraph matching algorithms \cite{topkdag, 73}. Typically, $\alpha$-nDCG, P-Coverage, Jaccard Distance, M-IA, etc. \cite{diverseS} are used to evaluate the diversity of results in information retrieval and recommendation systems. For database systems, there are also result diversity rules \cite{74, 76}, such as tuple coverage and max-min distance. Some works take the diversity of subgraph matching results into consideration \cite{topkdag, 73}, for example, DSQL \cite{73} utilizes vertex coverage to assess the diversity. However, none evaluate results using topological metrics.

\section{Conclusion} \label{sec:7}
We study the problem of top $k$ diversified subgraph matching and identify limitations in existing methods in diversity quality. To address this, we propose a distance-based diversity metric and a partition-based framework with two optimizations that improve scalability and result dispersion. As future work, we will also explore incorporating richer topological structures of the graph to further enhance our method.
%

\section*{ACKNOWLEDGEMENT} 
This work is supported by the Creative Research Groups Program of the National Natural Science Foundation of China (Grant No. 62321003), the National Natural Science Foundation of China (Grant Nos. U23A20317, 62572182, 62402481, 62576051), and the Natural Science Foundation of Hunan Province (Grant Nos. 2023JJ10016, 2023JJ30083).

\newpage
\balance

\bibliographystyle{ACM-Reference-Format}
\bibliography{reference}
\newpage
\appendices 
\section*{Appendix}
\vspace{1mm} 
\section*{A. Theoretical Analysis}

\setcounter{theorem}{0}
\begin{theorem}
Finding the top-$k$ subgraph matches that maximize distance-based diversity is NP-Hard, even when all subgraph matches are pre-computed.
\end{theorem}

\begin{proof}
    We assume that all subgraph matches of query $Q$ in data graph $G$ have already been computed, forming the candidate set $\mathcal{R}$. We reduce from the classical \emph{Max–Min $k$-Dispersion} problem,
which is known to be NP-hard on metric inputs.
\noindent\underline{Max–Min $k$-Dispersion.}
Given a finite point set $P = \{p_1, \dots, p_n\}$ and a distance function
$d: P \times P \to \mathbb{R}_{\ge 0}$ that is non-negative, symmetric,
and satisfies $d(p,p)=0$ (and possibly the triangle inequality), and a parameter $k \le n$, the goal is to find a subset
\begin{equation}
    \setlength\abovedisplayskip{1pt}
    \setlength\belowdisplayskip{1pt}
    S^* \subseteq P,\quad |S^*| = k,
    \text{that maximizes} 
    \min_{u \ne v \in S^*} d(u, v).
\end{equation}
\noindent\underline{Reduction construction.}
Let $(P, d, k)$ be an arbitrary instance of Max–Min Dispersion.
\begin{itemize}[leftmargin=4pt, nosep]
    \item Data graph. For each point $p_i \in P$, create a vertex $v_i$.
    Construct a complete undirected graph $G = (V,E)$ over $\{v_1, \dots, v_n\}$,
    and set the weight of each edge $(v_i, v_j)$ to $d(p_i, p_j)$.
    Because $d$ satisfies the triangle inequality, the shortest-path
    distance between $v_i$ and $v_j$ in $G$ equals $d(p_i, p_j)$.
    \item Query and matches. Let $Q$ be a single isolated vertex. Every
    $v_i$ is a valid match. Define the candidate set as
    $\mathcal{R} = \{ \{v_1\}, \dots, \{v_n\} \}$.
    \item Distance between matches. For any two matches
    $R_i = \{v_i\}$ and $R_j = \{v_j\}$, define the inter-match distance as
    \begin{equation}
        \setlength\abovedisplayskip{1pt}
        \setlength\belowdisplayskip{1pt}
        d^*(R_i, R_j) = \operatorname{dist}_G(v_i, v_j) = d(p_i, p_j).
    \end{equation}
    \item Objective. The DT$k$SM problem asks to find a subset
$S \subseteq \mathcal{R}$ of size $k$ that maximizes the minimum
inter-match distance:
\begin{equation}
    \setlength\abovedisplayskip{0pt}
    \setlength\belowdisplayskip{1pt}
    \max_{S \subseteq \mathcal{R},\,|S|=k}
    \min_{R_i \ne R_j \in S} d^*(R_i, R_j).
\end{equation}
\end{itemize}
\noindent\underline{Equivalence.}
Any solution $S \subseteq \mathcal{R}$ corresponds to a subset
$P_S \subseteq P$ of the same size, and since $d^*(R_i, R_j) = d(p_i, p_j)$,
the two problems are equivalent. \\
\noindent\underline{Complexity.}
The transformation uses $O(n^2)$ edges and $n$ candidate matches,
so its time and space costs are polynomial in $n$. Thus
Max–Min Dispersion $\le_P$ DT$k$SM, proving that DT$k$SM is NP-hard.

\end{proof}

Under the specific assumptions, the approximation ratio $\rho$ of the distance-based diversity is as follows:

\paragraph{Assumptions (A1)--(A5).}
Throughout, we consider a graph partitioned into disjoint subgraphs (or “partitions”).  We impose the following conditions on each partition \(G_k\):
\begin{enumerate}
  \item[(A1)] \emph{Compactness.}  Each partition \(G_k\) lies within a ball of fixed radius \(d\) around its center \(c_k\), i.e.,
  \[
  \max_{x \in G_k}\mathrm{dist}_G(x, c_k) \;\le\; d.
  \]
  \item[(A2)] \emph{Uniformity.}  The radius \(d\) in (A1) is the same for all partitions.
  \item[(A3)] \emph{Boundary thickness.}  When moving from one partition to the next along a path in the Partition Adjacency Graph (PAG), any path that crosses an internal boundary must travel at least \(2d\).
  \item[(A4)] \emph{Filtering success.}  After a filtering stage, every retained partition contains at least one valid match; that is, partitions selected by the algorithm are not dead ends.
  \item[(A5)] \emph{Match in the partition.} 
For analysis, we only consider matches that are fully contained within a single partition.
\end{enumerate}
\setcounter{definition}{0}
\begin{definition}[Approximation ratio]\label{def:approx_ratio}
For distance-based diversity, the approximation ratio is defined by
\[
\rho \;=\; \frac{D_{\mathrm{alg}}}{D^*},
\]
where \(D_{\mathrm{alg}}\) is the distance-based diversity achieved by the algorithm and \(D^*\) is the optimal distance-based diversity.
\end{definition}
\setcounter{lemma}{0}
\begin{lemma}[Distance envelope]\label{lem:envelopes}
Let \(G_i\) and \(G_j\) be partitions whose hop distance in the PAG is \(h=d_H(G_i,G_j)\).  Under (A1) and (A2), the true distance between any vertices \(u\in G_i\) and \(v\in G_j\) satisfies
\[
2\,(h-1)\,d \;\le\; \mathrm{dist}_G(u,v) \;\le\; 2\,(h+1)\,d.
\]
\end{lemma}

\begin{theorem}[Approximation ratio]\label{the:1}
Let \(G_i\) and \(G_j\) be the pair of partitions selected by the algorithm as the farthest in the PAG, with hop distance \(h=d_H(G_i,G_j)\).  Under assumptions (A1)–(A4) and the boundary thickness (A3) for the lower bound, the approximation ratio defined in Definition~\ref{def:approx_ratio} satisfies
\[
\frac{h-1}{h+1} \;\le\; \rho \;\le\; 1.
\]
\end{theorem}

\begin{proof}
Since the ratio is defined by \(\rho = D_{\mathrm{alg}}/D^*\), it is immediate that \(\rho \le 1\), because \(D_{\mathrm{alg}}\) cannot exceed \(D^*\).

For the lower bound, consider the partitions \(G_i\) and \(G_j\) with hop distance \(h\).  Lemma~\ref{lem:envelopes} gives
\[
2(h-1)d \;\le\; \mathrm{dist}_G(u,v)\;\le\; 2(h+1)d
\]
for \(u\in G_i\), \(v\in G_j\).  The algorithm selects a pair that realizes the left-hand (worst-case) bound, so its achieved diversity is \(D_{\mathrm{alg}} \ge 2(h-1)d\).  On the other hand, an optimal solution cannot exceed the right-hand bound, hence \(D^*\le 2(h+1)d\).  Therefore
\[
\rho \;=\; \frac{D_{\mathrm{alg}}}{D^*}
\;\ge\;
\frac{2(h-1)d}{2(h+1)d}
\;=\;
\frac{h-1}{h+1},
\]
proving the stated bounds on \(\rho\).
\end{proof}

The lower bound \(\tfrac{h-1}{h+1}\) increases with \(h\) and approaches \(1\) as \(h\to\infty\), meaning that for partitions far apart in the PAG, the approximation ratio becomes arbitrarily close to one.

\begin{remark}
Firstly, although Theorem~\ref{the:1} is stated for pairwise distances, it naturally extends to the top-$k$ setting. Specifically, every pair among the $k$ selected matches must satisfy the same lower bound. In addition, since we do not select adjacent partitions, the $h$-hop distance between any two chosen partitions in the PAG is always at least 2.
Furthermore, when the partition size decreases, the number of partitions increases, which makes it easier to select non-adjacent partitions that are farther apart in the PAG. As a result, the minimum inter-partition hop $h$ tends to grow, thereby improving the lower bound of distance diversity (cf. Theorem~\ref{the:1}). However, the partition size cannot be arbitrarily small. As illustrated in Experiment~1, when the partition size falls below the range of 1000–2000 vertices, the execution time increases because boundary vertex replication grows, leading to more cross-partition matches. Hence, there exists a clear trade-off between distance diversity (which benefits from larger $h$) and execution time (which deteriorates when the partition size is too small).
\end{remark}
\begin{figure*}[!htbp]
\vspace{-16pt}
    \centering
\includegraphics[width=1.0\linewidth]{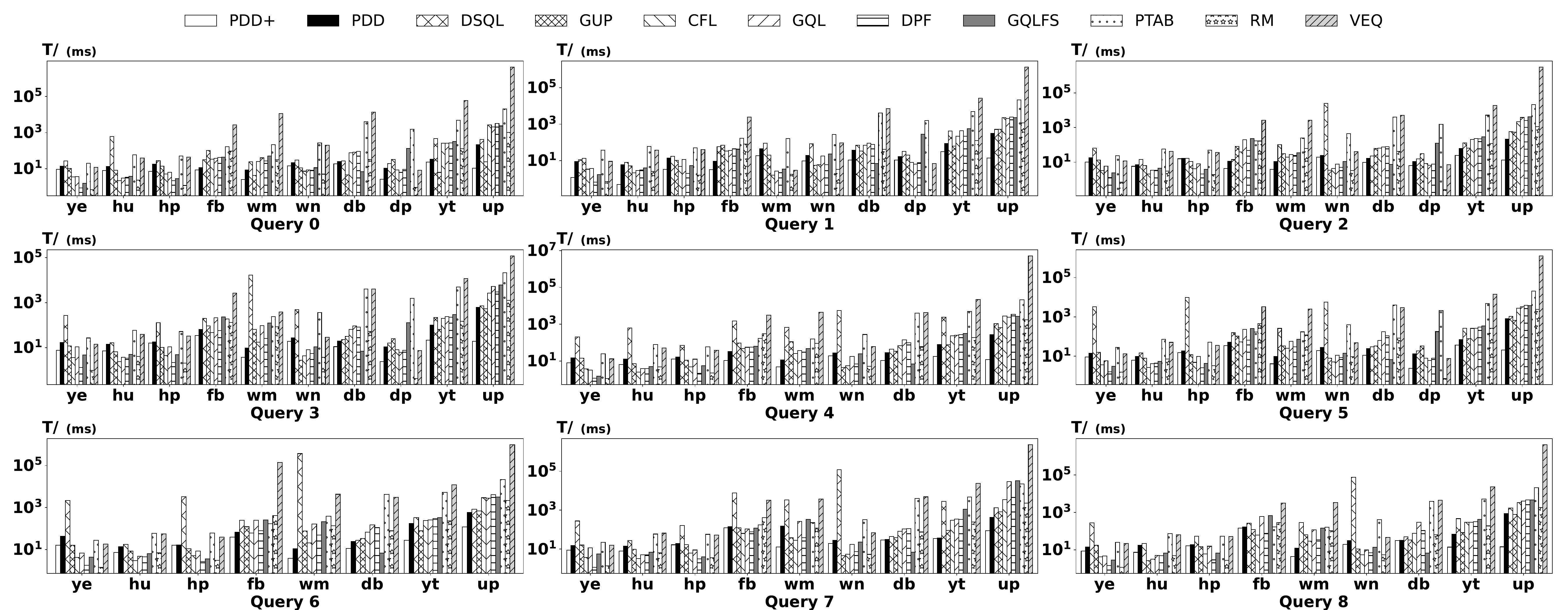}
    \caption{Runtime of Different Query on Different Real-World Datasets (\textit{k} = 30)}
    \label{fig:runtime}
\end{figure*}
\begin{figure}[!htbp]
    \centering
    \includegraphics[width=0.5\textwidth]{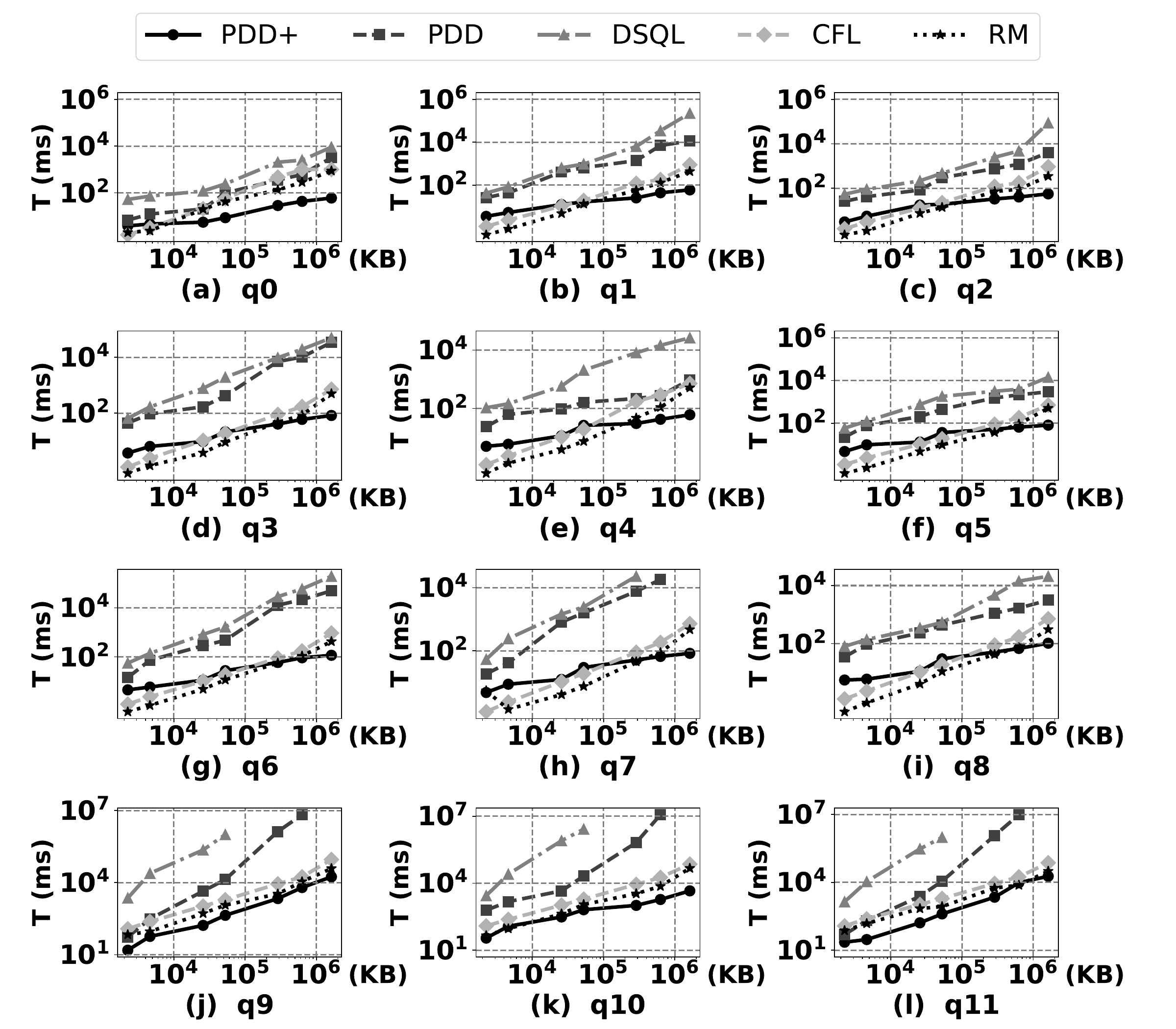}
    \caption{Runtime Variation with Data Size (k = 30).}
    \label{fig:k-time}
\end{figure}
\section*{B. Additional Experimental Results}

As shown in Figure~\ref{fig:runtime}, we evaluate PDD+ and baselines across all queries and real-world datasets. Some datasets are omitted for some queries when no valid match exists.
%
For simple queries (Query 0-3), PDD+ generally achieve lower runtimes across most datasets, with a clear advantage on large graphs. As dataset size increases, the runtime of most baselines escalates rapidly, whereas PDD and PDD+ remain efficient. PDD+ also consistently outperforms DSQL, reducing latency by approximately 70\%. On smaller datasets, subgraph enumeration algorithms sometimes outperform PDD+; this is attributable to the comparable sizes of the partition graph and the original graph. On larger datasets, PDD+ again surpasses these classical approaches.
For moderately complex queries (Query 4-8), PDD+ achieves over an order-of-magnitude speed-up compared to DSQL, highlighting its robustness as query complexity increases.
For near-clique queries (Query 9–11), PDD+ still achieves two to three orders of magnitude speed-up over DSQL, consistently shows clear advantages on medium and large datasets. \\
\begin{figure}[!htbp]
    \centering
    \includegraphics[width=0.5\textwidth]{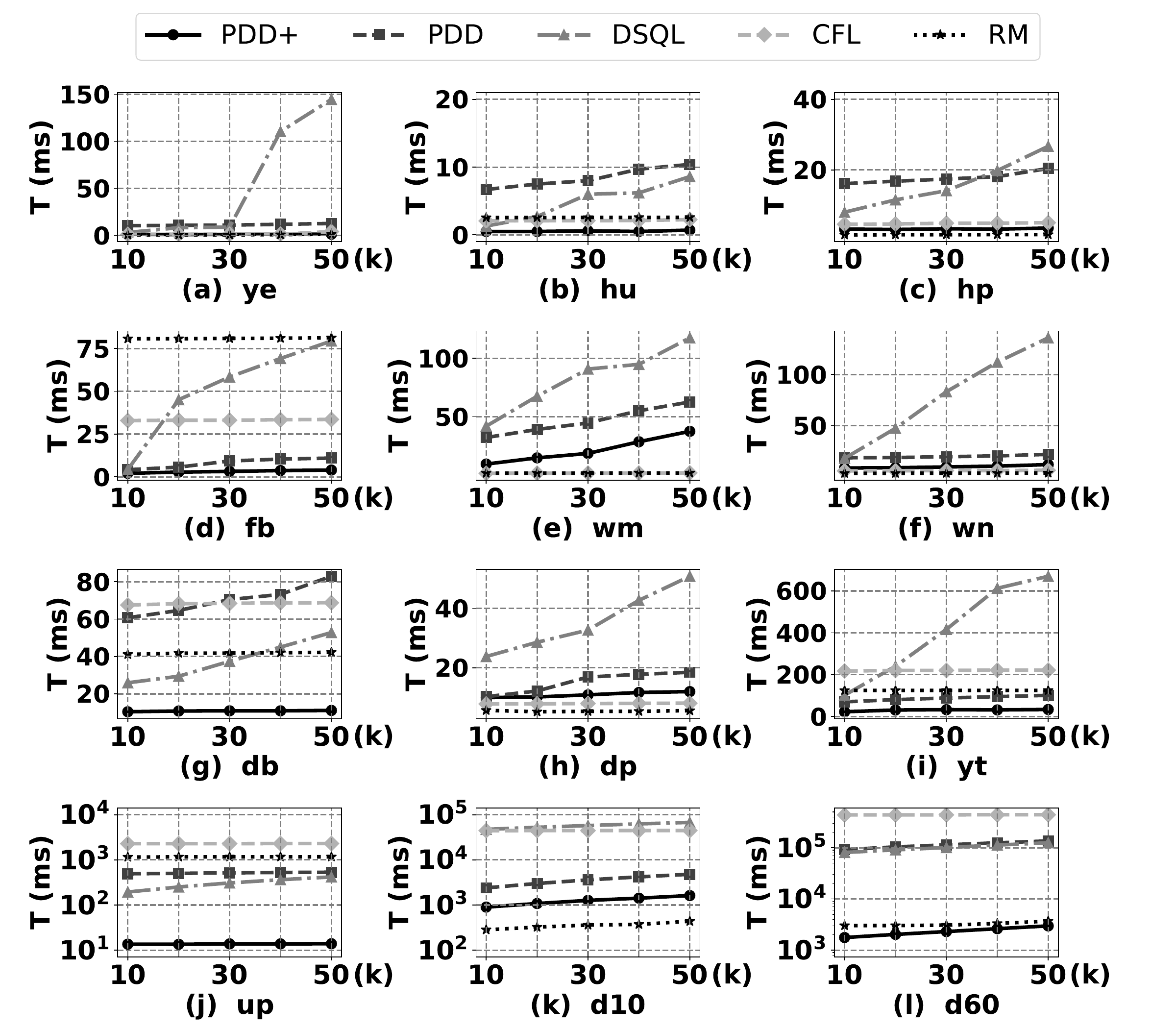}
    \caption{Runtime Variation with \textit{k}.}
    \label{fig:k}
\end{figure} \\
As shown in Figure \ref{fig:k-time}, for simple queries, PDD+ consistently performs the fastest, while DSQL is noticeably slower and RM and CFL fall in between. As query complexity increases, PDD+ shows the slowest growth in runtime, whereas DSQL rises exponentially. For complex queries, the runtime of all methods increases by several orders of magnitude, yet PDD+ still maintains the lowest curve and the smoothest growth, indicating strong robustness and scalability.

Figure \ref{fig:k} presents the runtime comparison of different methods across 12 real-world datasets, where $k$ varies from 10 to 50. Overall, PDD+ consistently achieves the lowest runtime on all datasets, demonstrating excellent scalability and stability. For smaller datasets, PDD+ still maintains a clear advantage. As the dataset size increases, the performance gap widens—DSQL shows a steep growth in runtime, while PDD+ and PDD exhibit smoother and much lower growth trends. On the largest datasets, DSQL and CFL experience exponential increases in runtime, whereas PDD+ maintains near-linear scalability.

\end{document}